\documentclass[aps,prd,onecolumn,showpacs,groupedaddress]{revtex4-2}
\usepackage{graphicx}
\usepackage{dcolumn}
\usepackage{bm}
\usepackage{color}
\usepackage{longtable}
\usepackage{supertabular}
\usepackage[T1]{fontenc}
\usepackage{epstopdf}
\usepackage{amsmath}
\usepackage{amssymb}
\usepackage{setspace}
\usepackage{multirow}
\usepackage{booktabs}
\usepackage[section]{placeins}
\usepackage{mathrsfs}
\usepackage{hyperref}

\makeatletter

\newcommand{\Rmnum}[1]{\expandafter\@slowromancap\romannumeral #1@} 
\newcommand{\bq}{\begin{equation}}
\newcommand{\eq}{\end{equation}}
\newcommand{\bqn}{\begin{eqnarray}}
\newcommand{\eqn}{\end{eqnarray}}
\newcommand{\nb}{\nonumber}

\makeatother

\begin{document}

\bibliographystyle{unsrt}

\title{Onset of chaotic gravitational lensing in non-Kerr rotating black holes with quadrupole mass moment}

\author{Wen-Hao Wu$^{1}$}
\author{Cheng-Yong Zhang$^{2}$}
\author{Cheng-Gang Shao$^{1}$}\email[E-mail: ]{cgshao@hust.edu.cn}
\author{Wei-Liang Qian$^{3,4,5}$}\email[E-mail: ]{wlqian@usp.br}

\affiliation{$^{1}$ MOE Key Laboratory of Fundamental Physical Quantities Measurement, Hubei Key Laboratory of Gravitation and Quantum Physics, PGMF, and School of Physics, Huazhong University of Science and Technology, 430074, Wuhan, Hubei, China}
\affiliation{$^{2}$ Department of Physics and Siyuan Laboratory, Jinan University, 510632, Guangzhou, China}
\affiliation{$^{3}$ Escola de Engenharia de Lorena, Universidade de S\~ao Paulo, 12602-810, Lorena, SP, Brazil}
\affiliation{$^{4}$ Faculdade de Engenharia de Guaratinguet\'a, Universidade Estadual Paulista, 12516-410, Guaratinguet\'a, SP, Brazil}
\affiliation{$^{5}$ Center for Gravitation and Cosmology, School of Physical Science and Technology, Yangzhou University, 225002, Yangzhou, Jiangsu, China}

\date{Oct. 19th, 2022}

\begin{abstract} 
In the electromagnetic channel, chaotic gravitational lensing is a peculiar phenomenon in strong gravitational lensing. 
In this work, we analyze the properties and emergence of chaotic gravitational lensing in the Manko-Novikov black hole spacetime. 
Aiming to understand better the underlying physics, we elaborate on the boundaries of the accessible region in terms of the analyses of the contours of the effective potentials.
The latter is associated with the two roots of a quadratic equation.
In particular, we explore its interplay with ergoregion, which leads to specific features of the effective potentials, such as the emergence of cuspy edge and the formation of {\it pocket}, that serves as a static constraint on the geodesics.
Besides, we investigate the properties of the radial and angular accelerations at the turning points in photons' trajectories.
Moreover, the accelerations are analyzed, which is argued to provide a kinematic constraint on the geodesics.
It is concluded that the onset of the chaotic lensing is crucially related to both constraints and as a result, an arbitrarily slight deviation in the incident photon is significantly amplified during the course of evolution through an extensive period, demonstrating the complexity in the highly nonlinear deterministic gravitational system.

\end{abstract}

\maketitle
\newpage

\section{Introduction}\label{section1}

Gravitational lensing is a prominent demonstration of Einstein's general relativity.
It is typically manifested in terms of the distortion of a galaxy's image in its weak form and the appearance of Einstein's ring in its strong form.
Although gravitational lensing was first proposed as a theoretical speculation~\cite{Eddington1922Space, Astronomische1924, Einstein1936}, its direct observation was not possible at the moment.
In the 1960s, the discovery of quasars~\cite{schmidt19633c} indicated that observing the phenomenon might be feasible, mainly because the quasars are excellent light sources in the universe due to their brightness.
Subsequently, weak gravitational lensing was first observed in 1979~\cite{walsh19790957+}, where the twin images of a distant quasar were published.
Besides the weak lensing effect, ultra-compact objects~\cite{Virbhadra:2022ybp} might bend light at more significant angles that, in turn, lead to a non-perturbative effect by creating even more extreme lensing images.
The latter is referred to as strong gravitational lensing~\cite{Kochanek2006}. 
Empirical observations unambiguously demonstrated that gravitational lensing is one of the most relevant observables in astrophysics. Since then, various aspects of weak gravitational lensing effects have been reported further, including numerous distinct gravitational lensing images and Einstein rings~\cite{jauncey1991unusually}. 

Moreover, in the strong field limit, a black hole might even cause light to revolve around it following given orbits.
For spherical black holes, the corresponding circular orbits form the light rings.
The latter is generalized to the notion of fundamental photon orbit (FPO)~\cite{Cunha2017fpo} by Cunha {\it et al.}
The FPO governs the boundary of the critical light geodesics that can be seen by an observer, known as the black hole shadow~\cite{agr-strong-lensing-shadow-review-01, agr-strong-lensing-shadow-review-02, agr-strong-lensing-shadow-review-03}.

In the case of the Kerr black hole, the relevant FPOs consist of spherical orbits.
Besides the structure of the spherical orbits determining the black hole shadow, the entire collection of light geodesics gives rise to the image of the black hole, a possible nearby accretion disk, on top of that of the background celestial sky.
The recent observation of the M87* supermassive black hole by the Event Horizon Telescope Collaboration~\cite{akiyama2019first} strongly indicated the significance of the electromagnetic channel playing an essential role in the novel era of precise astrophysics.
In many cases, the light geodesics, particularly the FPOs, possess a rich structure, giving rise to a sophisticated black hole image. 
In this regard, topics regarding black hole shadow and the image have aroused much interest in recent years~\cite{virbhadra2022distortions,Tang:2022hsu, Kuang:2022ojj, Tang:2022bcm, Sun:2023syd, Tang:2023lmr, Wang:2023vcv, Wu:2023wld, Meng:2023wgi,Ghorani:2023hkm,Xavier:2023exm}.
In particular, in Kerr black holes with Proca hair, the black hole shadow was demonstrated to possess a cuspy edge~\cite{Cunha2017fpo}, which has been analogically explained in terms of the Maxwell condition in the transition of a two-component system~\cite{qian2022cuspy}.
In the Randall-Sundrum braneworld scenario, a break in the shadow's boundary was observed, giving rise to an open shadow~\cite{hou2021revisiting}. 
Moreover, the lensing image of a Kerr black hole with scalar hair has been shown to lead to chaotic lensing~\cite{Cunha2016chaotic}.
In literature, chaotic scattering was known in the multibody scattering~\cite{shipley2016binary,yurtsever1995geometry}.
In the context of null geodesics in black hole spacetime, a few other spacetime structures were discovered to be subject to the phenomenon.
They include Bonnor black dihole~\cite{wang2018shadows} and a non-Kerr black hole with a quadrupole mass moment~\cite{wang2018chaotic}. 

The present paper further explores chaotic gravitational lensing while focusing on its generation mechanism.
We explore the lensing image of a non-Kerr black hole with a quadrupole mass moment.
The latter essentially is a Manko-Novikov spacetime with a single quadrupole deviation parameter. 
We investigate the effective potential of relevant geodesics and its connection to the emergence of chaotic lensing.
The remainder of the paper is organized as follows.
In the following section, we briefly revisit the black hole metric in question, where the effective potential and ergoregions are discussed.
Sec.~\ref{section3} is devoted to studying chaotic lensing in the black hole image and their connection with the pocket formed in the effective potential and acceleration of the geodesic at the turning points.
The concluding remarks are given in the last section.

\section{The properties of null geodesics and effective potentials}\label{section2}

\subsection{The equation of motion for null geodesics}

To investigate the strong gravitational lensing, inclusively the chaotic lensing phenomenon, in Manko-Novikov spacetimes~\cite{manko1992generalizations}, we start by exploring the properties of the null geodesics and the associated effective potential.
Here, we focus on only a particular subclass of the Manko-Novikov metric.
It is essentially determined by three parameters.
Namely, the mass $M$, spin $S$ of the black hole, and a dimensionless parameter $q$.
The parameter $q$ measures the deviation of the Manko-Novikov spacetime quadrupole mass moment from that of a Kerr one.
The metric in Boyer-Lindquist coordinates is given by
\begin{equation}
ds^2=-f(dt-\omega d\phi)^2+\frac{\rho^2 e^{2\gamma} dr^2}{f\Delta}+\frac{\rho^2 e^{2\gamma} d\theta^2}{f}+\frac{\Delta \sin^2\theta d\phi^2}{f},
\end{equation}
where the metric parameters $\rho$, $\Delta$, $f$, $\omega$, and $\gamma$ are defined as follows
\bqn
\rho^2 &=& (r-M)^2-k^2 \cos^2\theta, \nb\\
\Delta &=& (r-M)^2-k^2, \nb \\
k &=& M\frac{1-\alpha^2}{1+\alpha^2}, \nb\\
f &=& e^{2 \Psi} \frac{A}{B},\nb \\
\omega &=& 2ke^{-2\Psi} \frac{C}{A}-4k\frac{\alpha}{1-\alpha^2},\nb\\
e^{2\gamma} &=& e^{2\gamma'}\frac{A}{(x^2-1)(1-\alpha^2)^2},
\eqn
with
\bqn
\alpha &=& \frac{-M+\sqrt{M^2-(S/M)^2}}{(S/M)}, \nb\\
\beta &=& q\frac{M^3}{k^3}, \nb\\
A &=& (x^2-1)(1+ab)^2-(1-y^2)(b-a)^2,\nb\\
B &=& [(x+1)+(x-1)ab]^2+[(1+y)a+(1-y)b]^2,\nb\\
C &=& (x^2-1)(1+ab)[(b-a)-y(a+b)]+(1-y^2)(b-a)[(1+ab)+x(1-ab)],\nb\\
\Psi &=& \beta\frac{P_2}{L^3},\nb\\
a &=& -\alpha\exp\left[-2\beta\left(-1+\sum\limits_{\ell=0}^{2} \frac{(x-y)P_\ell}{L^{\ell+1}}\right)\right],\nb\\
b &=& \alpha\exp\left[2\beta\left(1+\sum\limits_{\ell=0}^{2} \frac{(-1)^{3-\ell}(x+y)P_\ell}{L^{\ell+1}}\right)\right],\nb\\
\gamma' &=& \ln\sqrt{\frac{x^2-1}{x^2-y^2}}+\frac{3\beta^2}{2L^6}(P_3^2-P_2^2)+\beta\left(-2+\sum\limits_{\ell=0}^{2}\frac{x-y+(-1)^{2-\ell}(x+y)}{L^{\ell+1}}P_\ell\right),\nb\\
P_\ell &=& P_\ell\left(\frac{xy}{L}\right), \nb\\
L &=& \sqrt{x^2+y^2-1},\nb\\
x &=& \frac{r-M}{k},\nb\\
y &=& \cos\theta .\label{metricPara2}
\eqn
and $P_\ell(z)$ are the $l$-order Legendre polynomials.
Subsequently, the black hole's outer horizon radius is $r_h=M+k$~\cite{manko1992generalizations,lukes2010observable,gair2008observable,bambi2011constraining,destounis2021gravitational1}.

The null geodesics of the black hole spacetime govern the trajectories of photons.
The latter can be expressed in terms of Hamilton's equation 
\bqn
\dot{x}^\mu &=&\frac{\partial \mathscr{H}}{\partial p_\mu}, \nb\\
\dot{p}^\mu &=&\frac{\partial \mathscr{H}}{\partial x_\mu}.\label{eosGeodesic}
\eqn
of the following Hamiltonian
\begin{equation}
\mathscr{H} =\frac{1}{2} p_\mu p_\nu g^{\mu \nu}=\frac{1}{2} (p^2_r g^{r r}+p^2_\theta g^{\theta\theta}+p^2_t g^{tt}+p^2_\phi g^{\phi\phi}+ 2p_\phi p_t g^{t\phi})=0.
\end{equation}
where one defines $E\equiv -p_t$ and $L \equiv p_\phi$.
The Hamiltonian can be divided into the sum of two parts.
The first one is the kinetic term
 \begin{equation}
K\equiv p^2_r g^{r r}+p^2_\theta g^{\theta\theta},
\end{equation}
and the second term gives the potential energy
\begin{equation}
V\equiv p^2_t g^{t t}+p^2_\phi g^{\phi\phi}+2p_\phi p_t g^{t\phi}.\label{Veff}
\end{equation}
Since the kinetic energy is positive semi-definite, the potential term thus delimits the accessible region in the $(r,\theta)$ coordinates.

To be specific, Eq.~\eqref{Veff} can be rewritten as 
\begin{equation}
V=-\frac{1}{D}(E^2 g_{\phi\phi}+2ELg_{t \phi}+L^2 g_{tt})\le 0,
\end{equation}
where $D\equiv g_{t\phi}^2-g_{tt} g_{\phi \phi}>0$ outside the horizon.
By defining the impact parameters $\eta \equiv \frac{L}{E}$, it is convenient to rescale the effective potential to read
\begin{equation}
\bar{V}\equiv -\frac{D V}{E^2}=g_{\phi\phi}+2\eta g_{t \phi}+\eta^2 g_{tt} \ge 0,
\end{equation}
which can be further rewritten as
\begin{equation}
\bar{V}=g_{tt}(\eta-h_+)(\eta-h_-) \ge 0,\label{conVeff}
\end{equation}
where
\begin{equation}
h_\pm \equiv  \frac{-g_{t \phi}\pm\sqrt{D}}{g_{tt}}.\label{effHpm}
\end{equation}
It is noted that the condition Eq.~\eqref{conVeff} introduces two two-dimensional functions $h_\pm = h_\pm(r,\theta)$. 
Since $\bar{V}=0$, when $\eta=h_\pm$, the contours of $h_\pm(r, \theta)$, therefore, govern the boundary of the accessible region in the $(r, \theta)$ coordinates for a subclass of null geodesics, whose $\eta$ is given.
As discussed below, a transition occurs at the ergoregion boundary where the sign of $g_{tt}$ flips.
The specific solution of Eq.~\eqref{conVeff} will be elaborated further in the following subsections.

\subsection{The ergoregion and effective potentials}\label{Ergoreg}

An ergosurface~\cite{herdeiro2014ergosurfaces,herdeiro2016spinning} is constituted by a collection of spacetime where the timelike killing vector field (uniquely recognized at spatial infinity) becomes null. 
The ergoregion~\cite{comins1978ergoregion,jacobson1998event}, defined as the spacetime region between the outer horizon and outer ergosurface, is a region where physical objects cannot remain stationary. 
In the present context, a close relationship exists between the ergoregion and the boundary of the allowable region by null geodesics, evaluated by employing the effective potential defined above. 

For the metric in question, the boundary of the ergoregion is determined by surface $g_{tt}=0$. 
This is because, for the only time killing vector at asymptotic spatial infinity $k$, we have $k^2=g_{tt}$.
By using the specific form of $g_{tt}$
\begin{equation}
 g_{tt}=-f=e^{2 \Psi} \frac{A}{B}=-e^{2\beta\frac{P_2}{L^3}}\frac{(x^2-1)(1+ab)^2-(1-y^2)(b-a)^2}{[(x+1)+(x-1)ab]^2+[(1+y)a+(1-y)b]^2} ,
\end{equation}
it is readily to verify $\lim\limits_{r\to\infty}g_{tt}=-1$, while at the horizon $r=r_h$ with $\theta=\frac{\pi}{2}$, thus $x=1$ and $y=0$, and $g_{tt}$ is manifestly positive.
Therefore, $g_{tt}$ must attain zero between the horizon and infinity on the equatorial plane. 
One may also analyze the ergoregion on the symmetric axis as follows.
First, as one approaches the pole in the zenith direction by taking the limit $r=r_h$ and $\theta\to 0^+$, which implies $x=1$ and $y\to 1^-$, it gives
\bqn
\lim\limits_{\theta \to 0^+}g_{tt}({r = r_h^+})=e^{2\beta\frac{P_2}{L^3}} \lim\limits_{y \to 1^-}\frac{(1-y^2) (b-a)^2}{4+4 a^2}= 0^+ .
\eqn
In other words, $g_{tt}> 0$ at the horizon as one moves away from the symmetry axis in the zenith direction, subsequently indicating the existence of the ergoregion.
However, if one approaches the outer horizon along the axis in the radial direction, by taking the limit $r\to r_h^+$ and $\theta=0$, which implies $x\to 1^+$ and $y=1$, one subsequently finds
\bqn
\lim\limits_{r \to r_h^+}g_{tt}({\theta = 0})=-e^{2\beta\frac{P_2}{L^3}} \lim\limits_{x \to 1^+}\frac{(x^2-1)(1+ab)^2}{4+4a^2}= 0^- .\label{limit2gtt}
\eqn
The latter indicates that the thickness of the ergoregion vanishes along the axis.
We note that it is largely similar to the case of the Kerr black hole.
In fact, when sitting on the horizon $r=r_h$, one always has $g_{tt}\ge 0$ for an arbitrary $\theta$, which indicates that the ergoregion is topologically connected.
The above results are also confirmed numerically, as shown below in Fig.~\ref{fIGgtt}.
For the current metric, it is observed that the ergoregion is relatively thin at certain zenith angles.

To proceed, we will make use of the properties that $g_{\phi\phi}>0$ and $g_{t\phi}<0$.
The former is also a physical requirement that there is no closed timelike geodesics, and these properties persist in accordance with most relevant axisymmetric metrics~\cite{Cunha2016chaotic}.

On the outside of the ergregion, namely, $g_{tt}<0$, the two-dimensional effective potential $h_\pm$ satisfies
\begin{equation}
h_+ < 0 < h_-.\label{gttNeg}
\end{equation}
For a given null geodesic, where $\eta$ is given, the condition Eq.~\eqref{conVeff} is met when
\bqn
h_+(r,\theta) \le \eta \le h_-(r,\theta). \label{cond1}
\eqn

Similarly, when $g_{tt}>0$, the null geodesic enters the ergoregion, and one has
\begin{equation}
0 < h_- < h_+.\label{gttPos}
\end{equation}
Now, the condition Eq.~\eqref{conVeff} is ensured once given that
\bqn
\eta &\le& h_-(r,\theta),\nb \\  
\mathrm{or}\ \ \ \eta &\ge& h_+(r,\theta).\label{cond2} 
\eqn
For both cases, equality is assumed at the boundary of the accessible region.
In other words, the boundary of the accessible region corresponds to either $\eta=h_-$ or $\eta=h_+$.

It is worth noting as one approaches the ergoregion from the outside, namely, $g_{tt}\to 0_-$.
It is apparent that one of the solution $h_+\to -\frac{2g_{t\phi}}{g_{tt}}\to -\infty$ which diverge, while $h_-\to -\frac{g_{\phi\phi}}{2g_{t\phi}}\equiv h_-^0$, which remains finite.
Therefore, the condition Eq.~\eqref{cond1} falls to
\bqn
\eta \le h_-^0. \label{cond1Limit}
\eqn
On the other hand, if one approaches the ergoregion from the inside, namely, $g_{tt}\to 0_+$, the effective potential diverges as $h_+\to +\infty$, while $h_-\to h_-^0$ and remains finite.
Subsequently, the condition Eq.~\eqref{cond2} simplifies, once more, to an expression identical to Eq.~\eqref{cond1Limit}.

In practice, the value of $\eta$ is given in the first place via its angular momentum and energy~\cite{Cunha2016chaotic}.
If $\eta>0$, the relevant bound of the accessible region depends on the sign of $g_{tt}$, and the above-derived condition can be rewritten as $Q_1=1$, where
\bqn
Q_1\equiv \mathrm{sgn}(g_{tt})\left[\mathrm{sgn}(\eta-h_+)+\mathrm{sgn}(h_--\eta)\right]+\mathrm{sgn}(-g_{tt})\mathrm{sgn}(h_--\eta). \label{cond3a}
\eqn
where the staircase function $\mathrm{sgn}$ evaluates the sign of a quantity,
\bqn
\mathrm{sgn}(x)= \left\{\begin{matrix}1&\ \ \mathrm{if}\ \ x\ge 0\\0&\ \ \mathrm{if}\ \ x <0\end{matrix} \right.. \label{sgnFun}
\eqn
On the other hand, if $\eta < 0$, the condition of accessibility can be expressed as $Q_2=1$, where
\bqn
Q_2\equiv \mathrm{sgn}(g_{tt})\mathrm{sgn}(h_--\eta)+\mathrm{sgn}(-g_{tt})\mathrm{sgn}(\eta-h_+)=\mathrm{sgn}(-g_{tt})\mathrm{sgn}(\eta-h_+), \label{cond3b}
\eqn
whose form simplifies due to Eq.~\eqref{gttPos}, thus the condition dictated by the first line of Eq.~\eqref{cond2} is always met.

By putting Eqs.~\eqref{cond3a} and~\eqref{cond3b} together, one may define the quantity
\bqn
Q\equiv \mathrm{sgn}(\eta)Q_1+\mathrm{sgn}(-\eta)Q_2. \label{cond3}
\eqn
The condition of accessibility corresponds to $Q=1$.

It is worth mentioning that a null geodesic that satisfies the second line of Eq.~\eqref{cond2} can be considered as a ``bound state''.
This is because one might argue that such a photon will never be able to escape to spatial infinity.
Consider a null geodesic with a given impact parameter $\eta_0$ initially propagates inside the ergoregion so that one has $g_{tt}>0$ and $\eta_0 > h_+$.
However, as one approaches the boundary of the ergoregion, $h_+$ increases and approaches $+\infty$, which eventually breaks the second line of Eq.~\eqref{cond2}.
Therefore, for such a photon, the spacetime region in the vicinity of the boundary of ergoregion is prohibited.
In other words, its trajectory is bound as an inaccessible region surrounds it.
In this regard, as $g_{tt}$ flips its sign, a transition occurs regarding the accessible region for a specific class of null geodesics, as observed in the last subsection.
In particular, the class of null geodesics associated with the second line of Eq.~\eqref{cond2} cannot appear outside the ergoregion, while the class associated with the first line of Eq.~\eqref{cond2} merges with those governed by Eq.~\eqref{cond1}, in accordance with Eq.~\eqref{cond1Limit}.
In the following subsection, we will see that the above properties of $h_+$ also dictate the feature of the associated contours, which gives rise to additional constraints on the motion of geodesics.

\subsection{Contour map of the effective potentials and accessible regions}

Now, we proceed to present the ergoregions, relevant null geodesics, and contours of the effective potential that constraint the null geodesics, based on the criterion in terms of the $h_+$ and $h_-$, elaborated in the preceding subsection.
In the following figures, we utilize dashed blue curves to represent the boundary associated with $h_-$, corresponding to the contours $h_-=\eta$, and dot-dashed red curves denote those associated with $h_+$.
The solid black curves indicate the boundaries of the ergoregion governed by the condition $g_{tt}=0$.
In what follows, we will focus on the interplay between the ergoregion and the contours of the effective potential. 
In particular, we explore the formation and properties of the pocket formed in the latter.
Without loss of generality, for our numerical calculations in the remainder of the paper, we utilize two sets of metric parameters: 
\bqn
\mathrm{set}\ &1&: \ \ S=0.2M^2,\ q=2,\ M=1, \nb\\
\mathrm{set}\ &2&: \ \ S=0.98M^2,\ q=8,\ M=1 .\label{metricPar}
\eqn

In particular, the second set of the metric parameter is in the vicinity of an extreme black hole.
It is featured by a more significant deviation of the quadrupole moment from the Kerr black hole.
It also brings novel characteristics to the resulting strong gravitational lensing, as discussed below.

For better visualization, we also introduce a compact radial coordinate $R\in [0, 1]$ defined as
\bqn
R=\frac{R^*}{1+R^*}, 
\eqn
where
\bqn
R^*=\sqrt{r^2-r_h^2} .
\eqn

\begin{figure}[htp]
\includegraphics[scale=0.35]{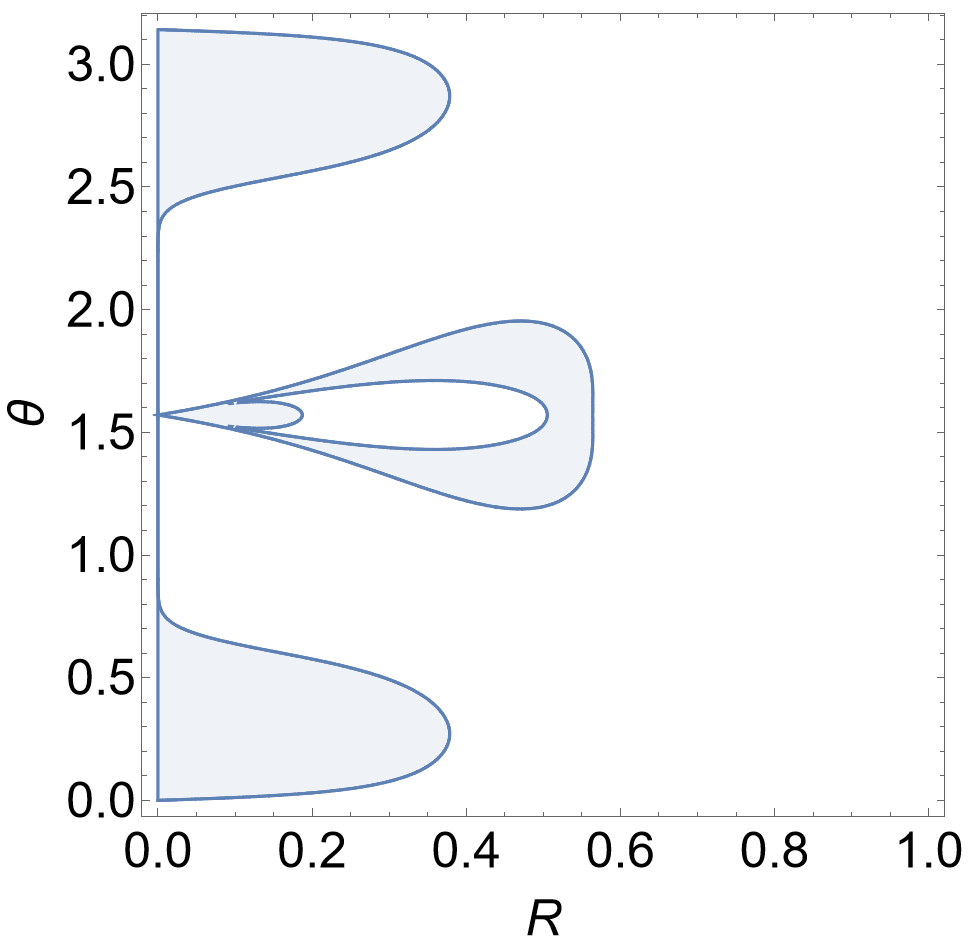}
\caption{The ergoregions of the specific Manko-Novikov black hole spacetime.}
\label{fIGgtt}
\end{figure}

In this subsection, the calculations are carried out using the first set of metric parameters.
We first present the ergoregions of the black hole metric in question in Fig.~\ref{fIGgtt}.
As discussed in the previous subsection, the ergoregion encloses the outer horizon $r=r_h^+$, which corresponds to the $y$-axis of the plot. 
At certain zenith angles, the thickness of the ergoregion is numerically insignificant in the present coordinate.
On the equatorial plane, one observes two separate areas of ergoregion, which will be referred to as outer and inner ergoregions.
Along the symmetry axis ($\theta=0$), though not visually apparent, the thickness of the ergoregion vanishes, in accordance with Eq.~\eqref{limit2gtt}.

\begin{figure}[htb]
\includegraphics[scale=0.28]{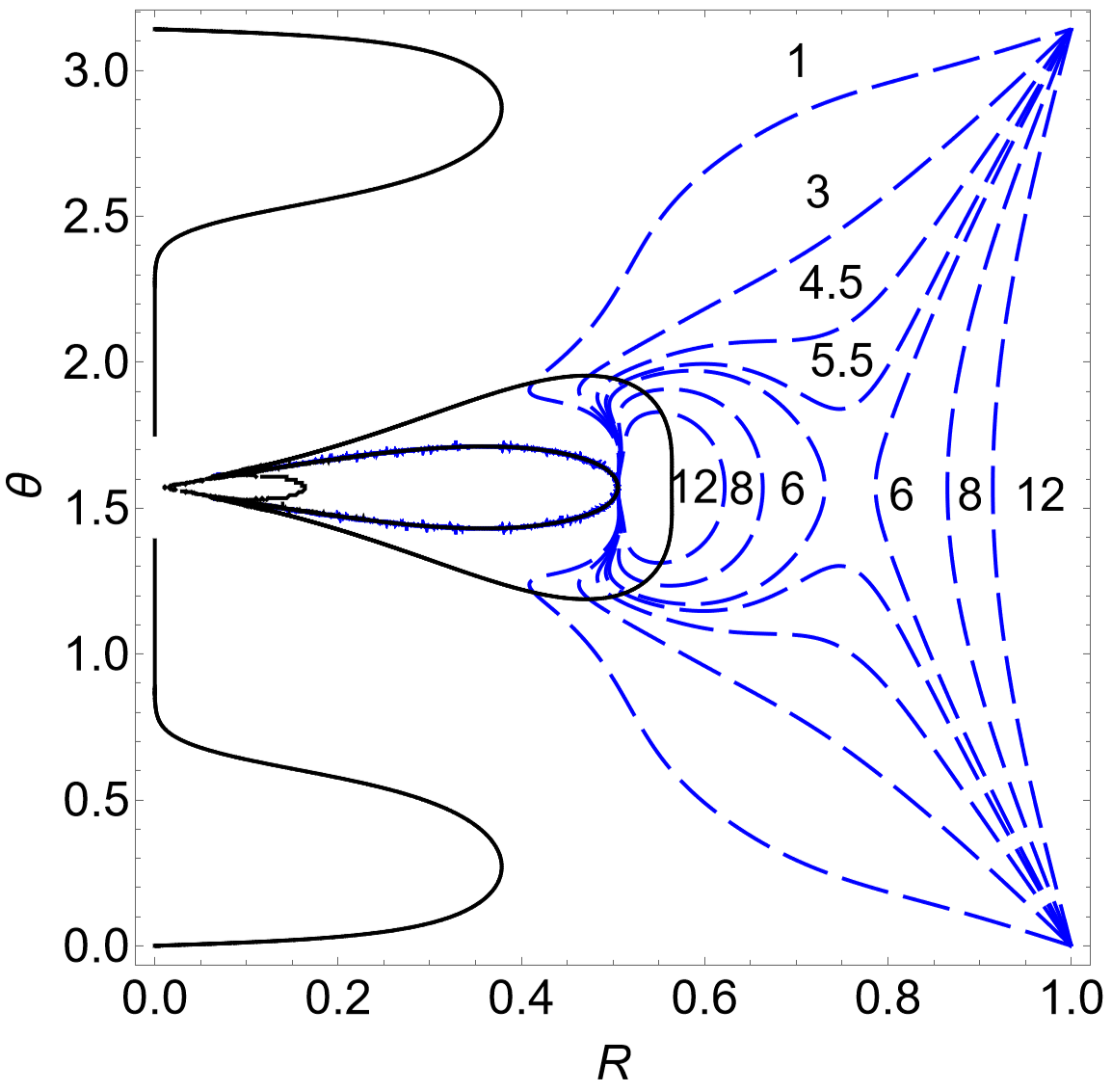}
\includegraphics[scale=0.34]{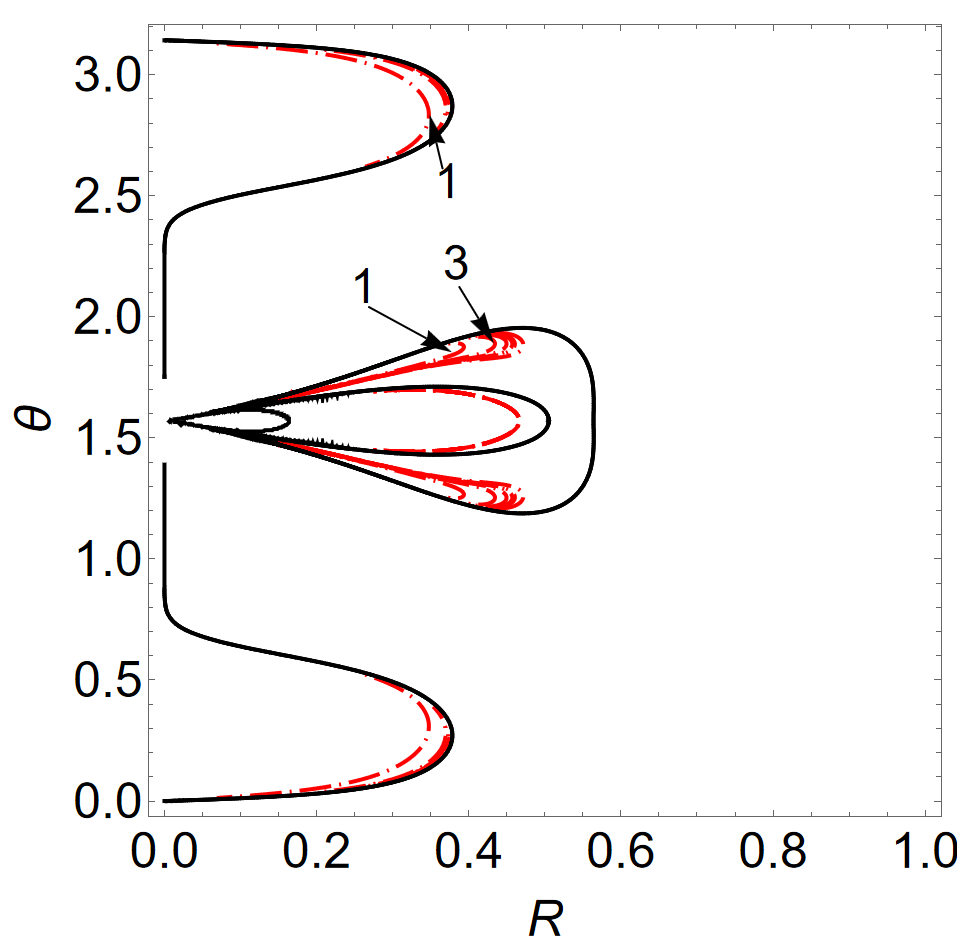}
\caption{ 
The contour maps $h_-=\eta$ (left) and $h_+=\eta$ (right) associated with the effective potentials Eq.~\eqref{effHpm} are shown, respectively, in dashed blue and dot-dashed red curves, where the values of $\eta\ (>0)$ are indicated.
The solid black curves indicate the boundary of the ergoregion.
}
\label{fig:figure1}
\end{figure}

In the left panel of Fig.~\ref{fig:figure1}, we present the contour map associated with the effective potential $h_-$.
Based on previous discussions regarding Eqs.~\eqref{cond1} and~\eqref{cond2}, the contours represent the boundary of the accessible region for null geodesics when the equalities are assumed.
Since $\eta>0$, these geodesics evolve around the black hole along the direction of its spin. 
On the right side of the contour map, one observes that the contours start to form pockets around $\eta\sim 6$ and even separate to form an isolated area of accessible region as $\eta$ further increases.
It is also noted that the blue dashed curves intersect with the solid black curves.
In other words, the union of these contours associated with $h_-$ corresponds to the geodesics capable of entering and exiting the ergoregion.
When examining more closely while comparing with Fig.~\ref{fIGgtt}, the contours cluster themselves in between the inner and outer ergoregion near the equatorial plane, as shown in the left panel of Fig.~\ref{fig:figure1}, they form a structure featured a cuspy edge. 
In particular, the blue dashed curves seem to break at the inner bound of the outer ergoregion.
Indeed, these counters enclose themselves around the immediate vicinity of the bound.
This is because $\eta> h_-^0 \sim -0.004$ around the inner boundary of the ergoregion, which is accompanied by a relatively steep gradient.
On the other hand, the accessibility of geodesics requires the condition $\eta< h_-$ given by the first line of Eq.~\eqref{cond2}.
Reminiscent of the discussions in the last subsection, such a geodesic will never be capable of traversing the bound since $h_-\to h_-^0 <\eta$ at the bound.
As a result, these contours that govern the boundary of the geodesics are constrained between the two ergoregions with a cuspy edge.
Similar to $h_-$, we also show the contour map of $h_+$ in the right panel of Fig.~\ref{fig:figure1}.
These contours stay mainly inside the ergoregion.
In the vicinity of the equatorial plane, they are located inside the outer ergoregion and outside the inner ergoregion.
According to the discussions in the last section, inside the ergoregion $g_{tt} >0$, and therefore, the geodesics constraint by such contours can never traverse the boundary of the outer ergoregion and escape to spatial infinity.

\begin{figure}[htb]
\includegraphics[scale=0.26]{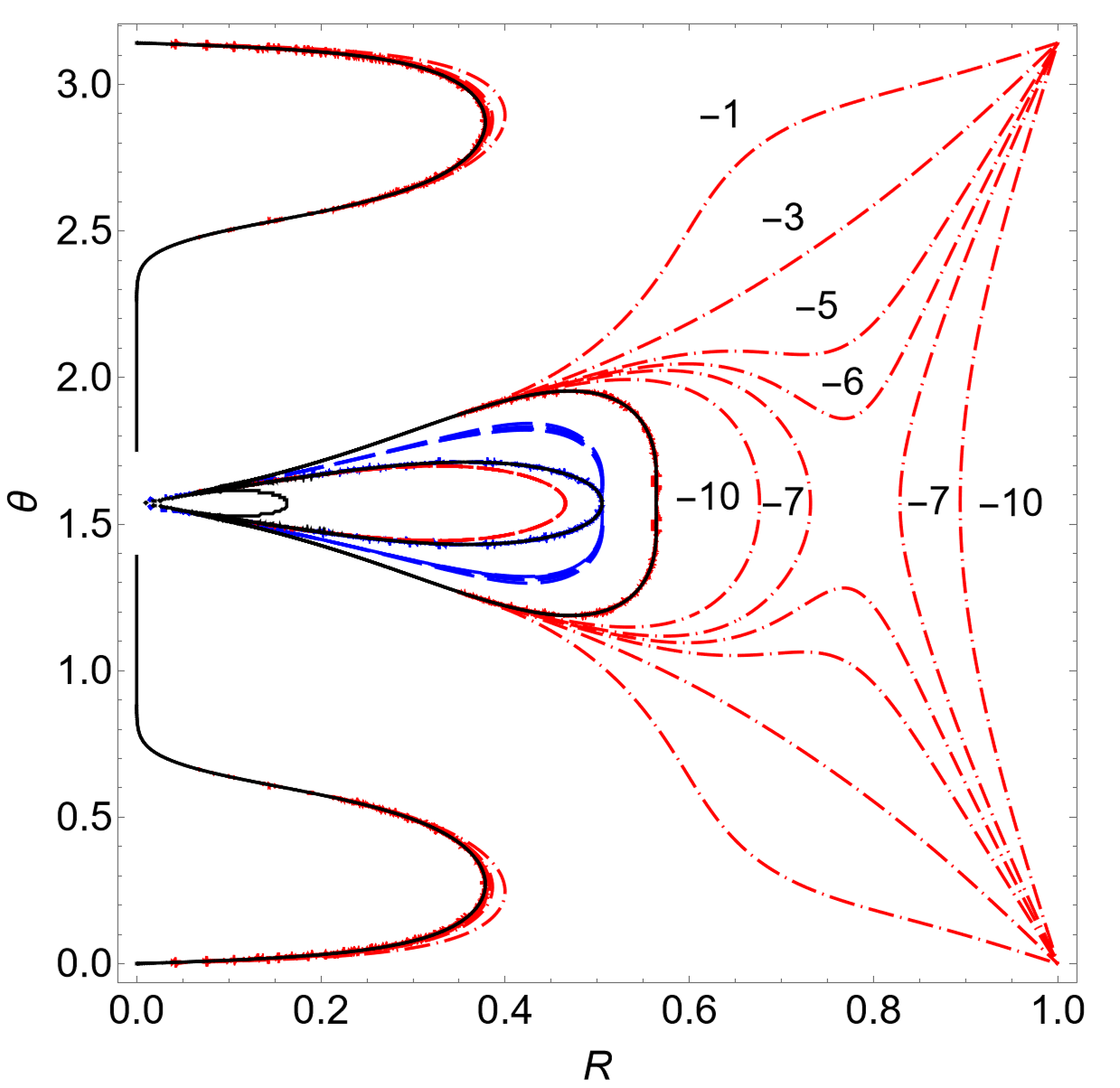}
\caption{
The same as Fig~\ref{fig:figure1} but for $\eta<0$, the contours associated with the effective potentials $h_-$ and $h_+$, shown in blue dashed and red dot-dashed curves.}
\label{fig:figure3}
\end{figure}

In Fig.~\ref{fig:figure3}, we present the contour map $h_\pm = \eta$ but for $\eta <0$.
These contours indicate possible constraints for geodesics evolving in the opposite direction of the black hole's spin.
It is apparent that the main features of the contours associated with the effective potential $h_\pm$ are exchanged.
Now, pockets also emerge by the contours related to $h_+$ as the value of $\eta \sim -7$.
As $\eta$ further decreases, the contour separates and forms isolated areas of accessible regions.
Different from the left panel of Fig.~\ref{fig:figure1}, the red dot-dashed curves do not traverse the boundary of the ergoregion.
This feature is observed for contours near the equator plane or the symmetry axis.
Such a result can also be derived by employing the arguments parallel to those given above.
Since $g_{tt}<0$ outside the ergoregion, we have $h_+ < 0$ according to Eq.~\eqref{gttNeg}.
As a geodesic approaches the bound from the outside, $h_+$ asymptotically approaches $-\infty$, which implies that an arbitrary geodesic with $h_+ <\eta <0$ will never be able to penetrate the ergoregion.
Subsequently, the pockets enclose themselves by marginally going through the boundary of the ergoregion.
In Fig.~\ref{fig:figure3}, the contours formed by the blue dashed curves are primarily located inside the outer ergoregion, similarly to the red dotted-dashed curves shown above in Fig.~\ref{fig:figure1}.


\begin{figure}[htb]
\includegraphics[scale=0.27]{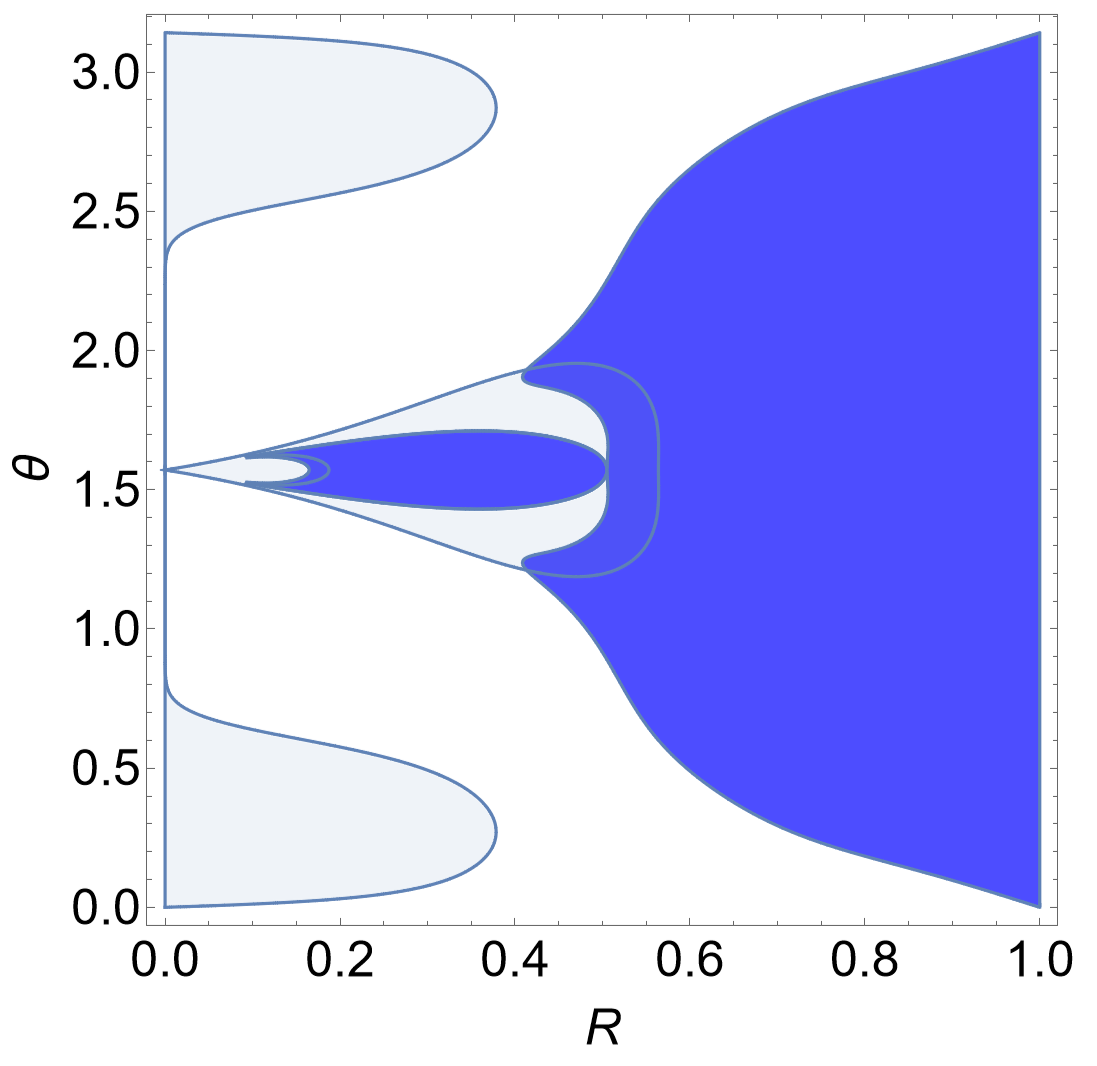}
\includegraphics[scale=0.25]{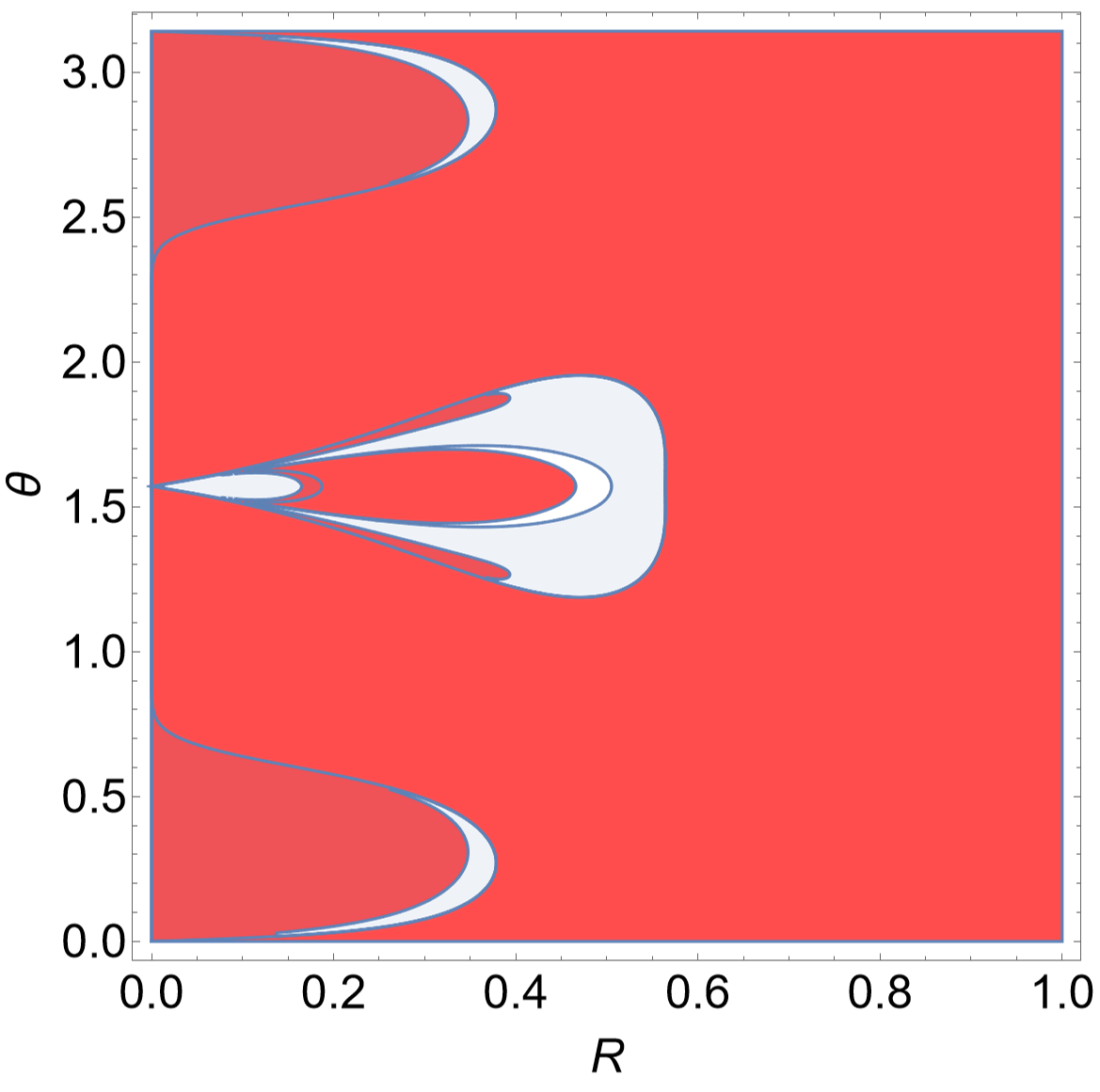}
\caption{The accessible region for geodesics with $\eta=1$ constraint by the effective potentials $h_-$ (left) and $h_+$ (right) based on Eq.~\eqref{cond3a}.}
\label{fig:h-eta-ergo}
\end{figure}

In what follows, we discuss the accessible region by combining the information of ergoregion and effective potential contours, as discussed in Eqs.~\eqref{cond3a}, ~\eqref{cond3b}, and~\eqref{cond3}.
In Fig.~\ref{fig:h-eta-ergo}, we show the accessible region for geodesics with $\eta=1$ based on Eq.~\eqref{cond3a}.
The left panel of Fig.~\ref{fig:h-eta-ergo} gives the region associated with the effective potential $h_-$, while the right panel represents that associated with $h_+$.
According to Eq.~\eqref{cond3a}, two contributions are related to the effective potential $h_-$.
They correspond to, respectively, the region outside the ergoregion where $g_{tt}<0$ and that inside the ergoregion with $g_{tt}>0$.
Since $\mathrm{sgn}(g_{tt})+\mathrm{sgn}(-g_{tt}) = 1$, the accessible region satisfies $\mathrm{sgn}(\eta-h_+) = 1$, irrelevant to $g_{tt}$.
As shown in the left panel of Fig.~\ref{fig:h-eta-ergo}, the accessible region is obtained by filling up the area enclosed by the contour $h_- =\tau =1$ and on the side where $\tau$ increases.
On the other hand, the accessible region associated with $h_+$ only concern the region inside the ergoregion, owing to the factor $\mathrm{sgn}(g_{tt})$ in Eq.~\eqref{cond3a}.
By comparing the right panel of Fig.~\ref{fig:h-eta-ergo} to that of Fig.~\ref{fig:figure1}, the above condition is evident as the fraction between the inner and outer ergoregions enclosed by the contour (in the dot-dashed red curve) in Fig.~\ref{fig:h-eta-ergo} is removed from the red area shown in Fig.~\ref{fig:h-eta-ergo}.

\begin{figure}[htb]
\includegraphics[scale=0.27]{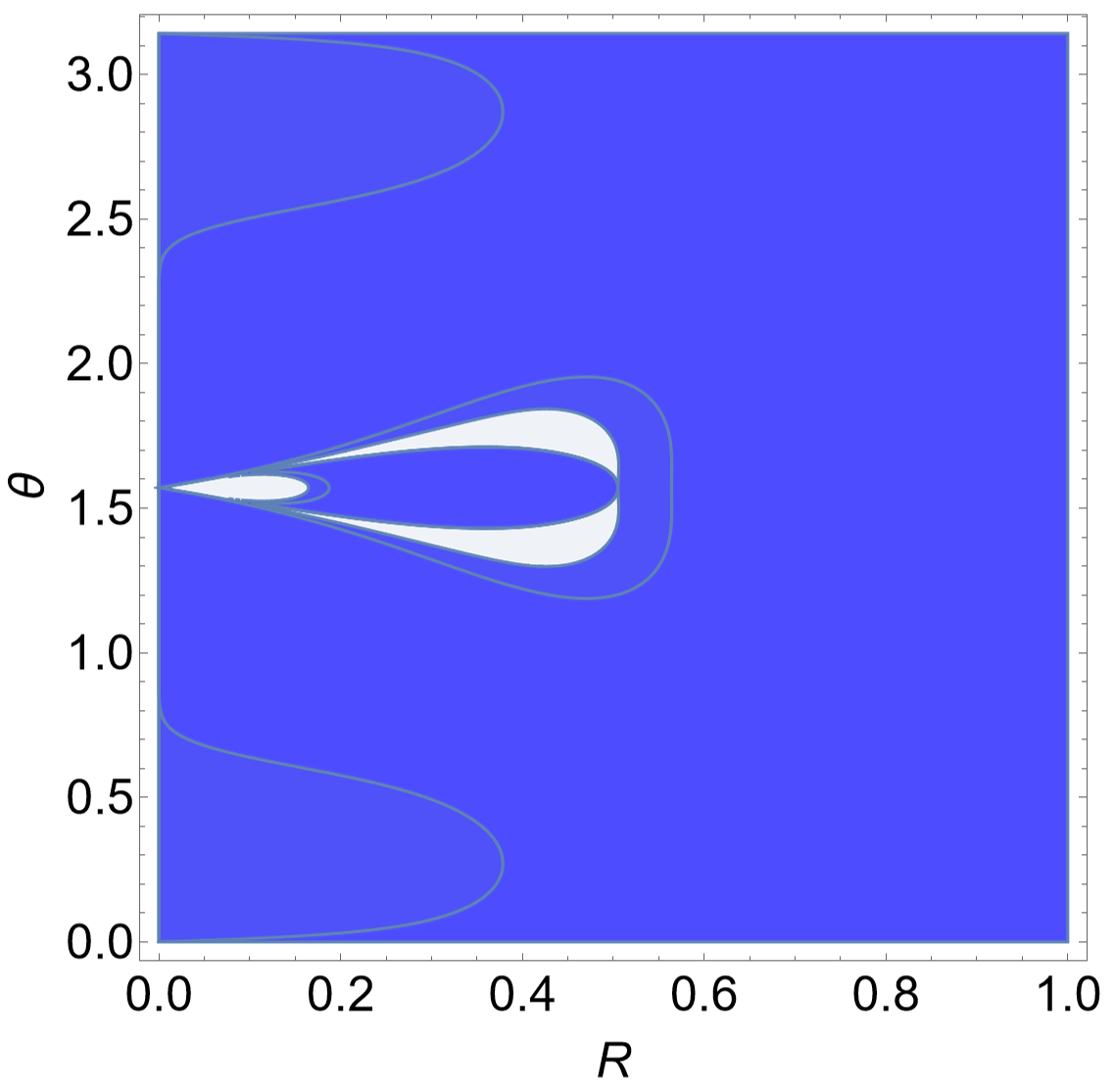}
\includegraphics[scale=0.28]{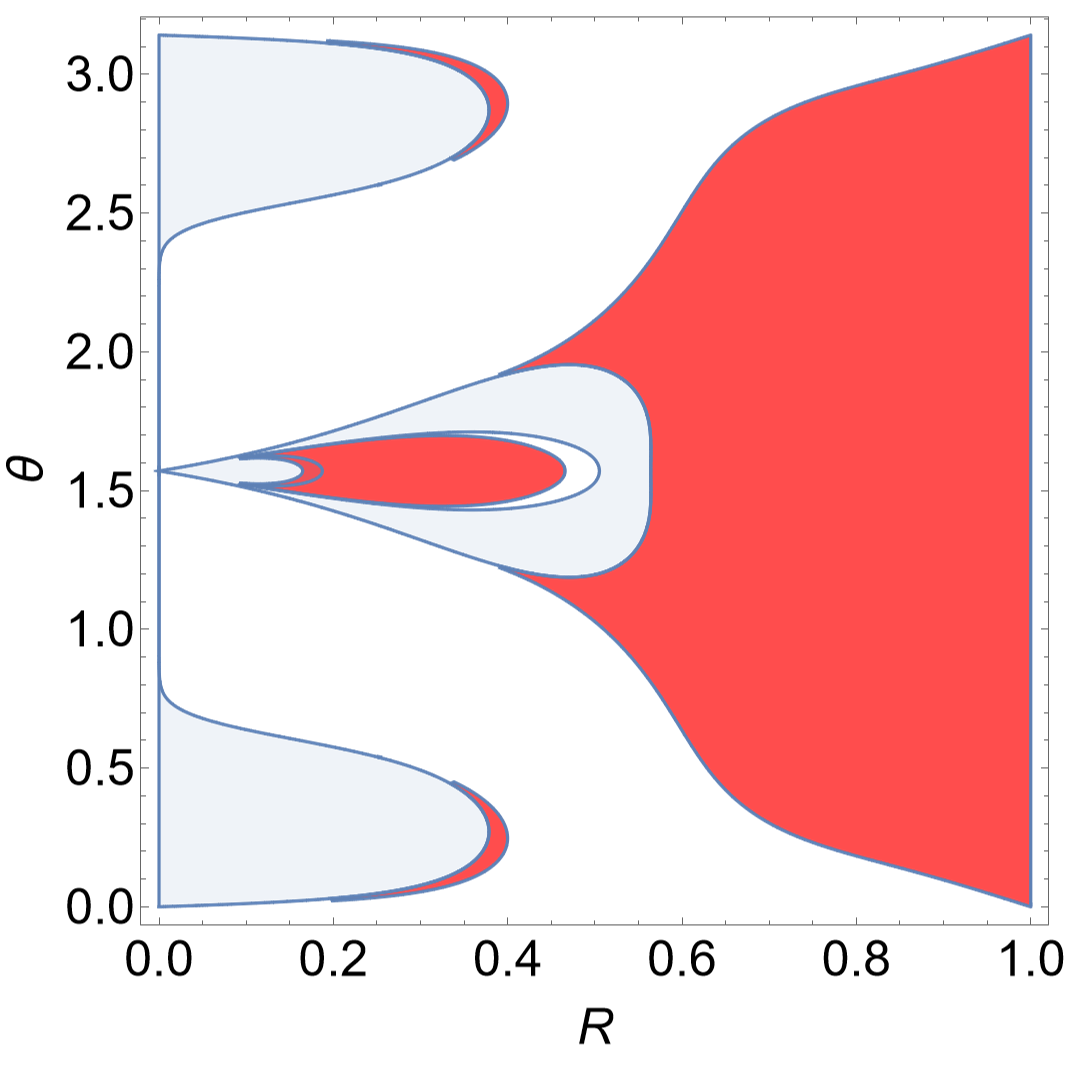}
\caption{The same as Fig~\ref{fig:h-eta-ergo} but for $\eta=-1$, the calculations are carried out for the effective potentials $h_-$ (left) and $h_+$ (right) based on Eq.~\eqref{cond3b}.}
\label{fig:h-eta-ergo2}
\end{figure}

Similarly, we present the accessible region in Fig.~\ref{fig:h-eta-ergo2} with with $\eta=-1 $ based on Eq.~\eqref{cond3b}.

\begin{figure}[htb]
\includegraphics[scale=0.31]{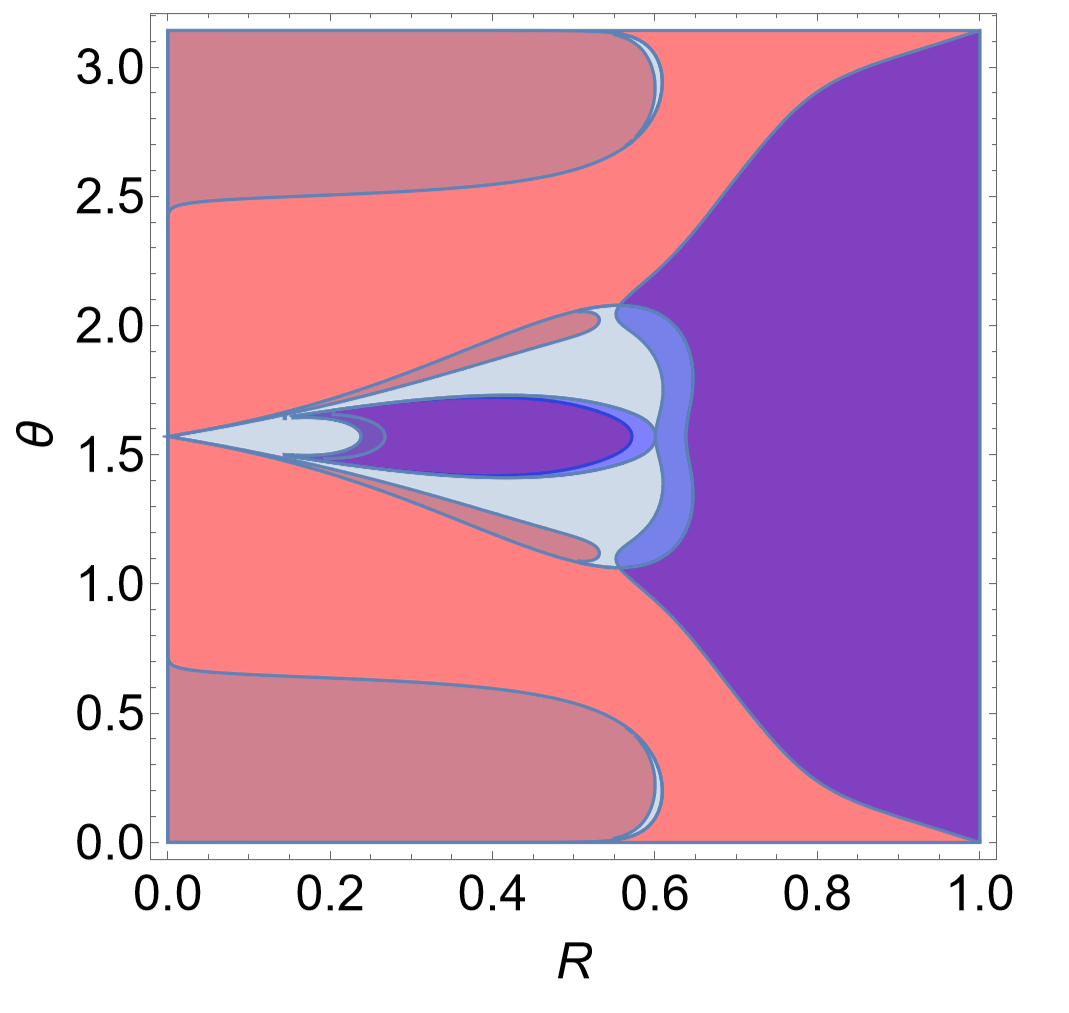}
\includegraphics[scale=0.29]{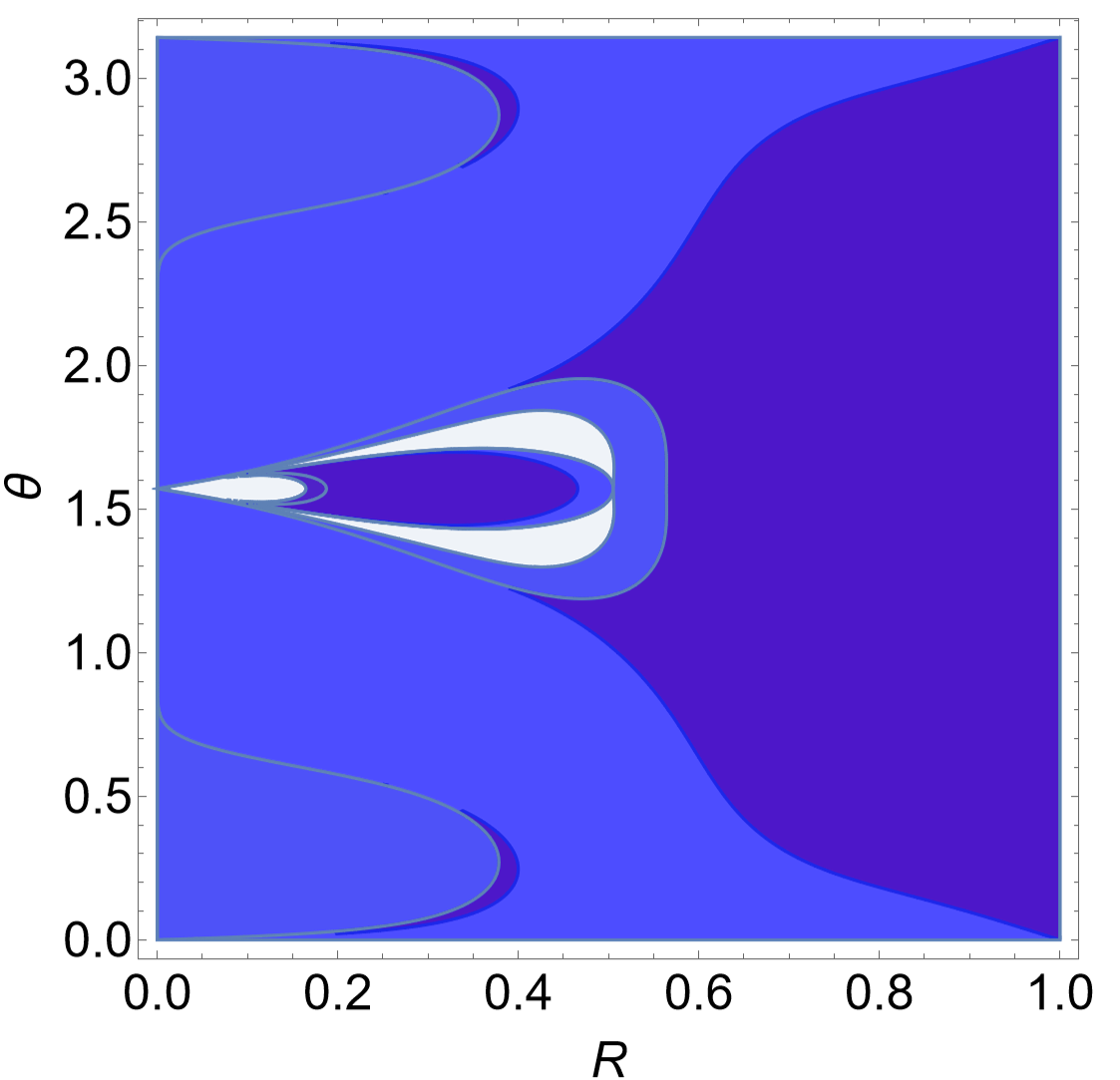}
\caption{The entire accessible regions for geodesics with $\eta=1$ (left) and $\eta=-1$ (right) based on Eq.~\eqref{cond3}.
The plots consist of the union of red ($h_+$) and blue ($h_-$) areas shown in Figs.~\ref{fig:h-eta-ergo} and~\ref{fig:h-eta-ergo2}, and it is filled by purple whenever the area is accessible by both effective potentials.}
\label{fig:h-eta-ergo3}
\end{figure}

Last but not least, by putting all the pieces together, the accessible region governed by Eq.~\eqref{cond3} is shown in Fig.~\ref{fig:h-eta-ergo3}.
These plots can be obtained by the unions or the intersection of Figs.~\ref{fig:h-eta-ergo} and~\ref{fig:h-eta-ergo2}.
Depending on the accessible regions, they are located inside or outside the ergoregions.
Alternatively, this can be obtained by directly employing Eq.~\eqref{cond3}.
We close this section by observing that some accessible regions are not topologically connected to spatial infinity but to the outer event horizon.
As a result, they are not physically relevant to the discussions of gravitational lensing.
The properties of the geodesics and accessible region explored in this subsection will be utilized to analyze the chaotic gravitational lensing observed in the black hole metric in question.

\section{Black hole images and chaotic gravitational lensing}\label{section3}

In this section, we first present the numerical results of black hole images in Manko-Noviko spacetime, featured by chaotic lensing, first reported by Wang, Mingzhi and Chen, Songbai and Jing, Jiliang~\cite{wang2018chaotic}.
Subsequently, we analyze the relations between the observed chaotic lensing and the properties of the geodesics explored in the preceding sections.

\subsection{Numerical results from backward ray-tracing calculations}

\begin{figure}[htb]
\includegraphics[scale=0.42]{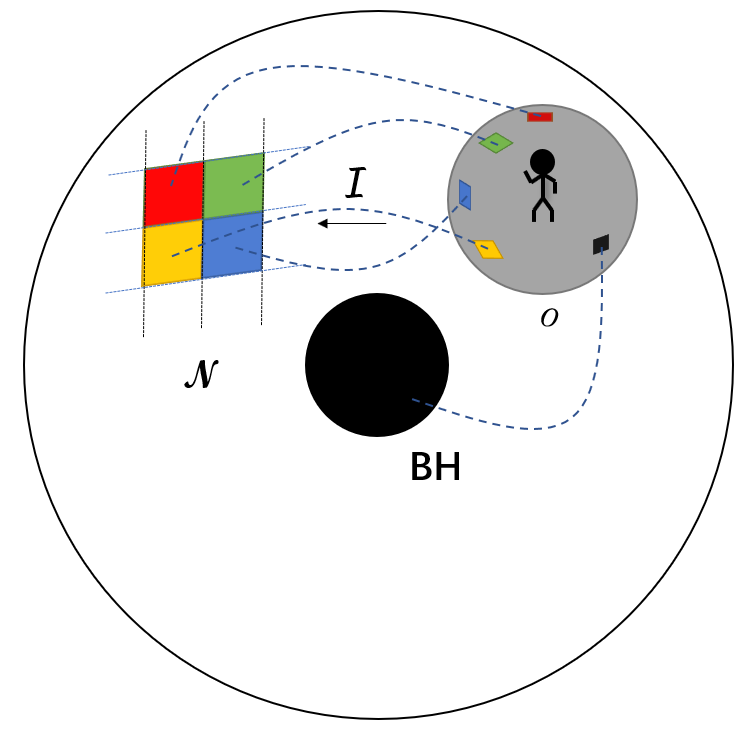}
\includegraphics[scale=0.26]{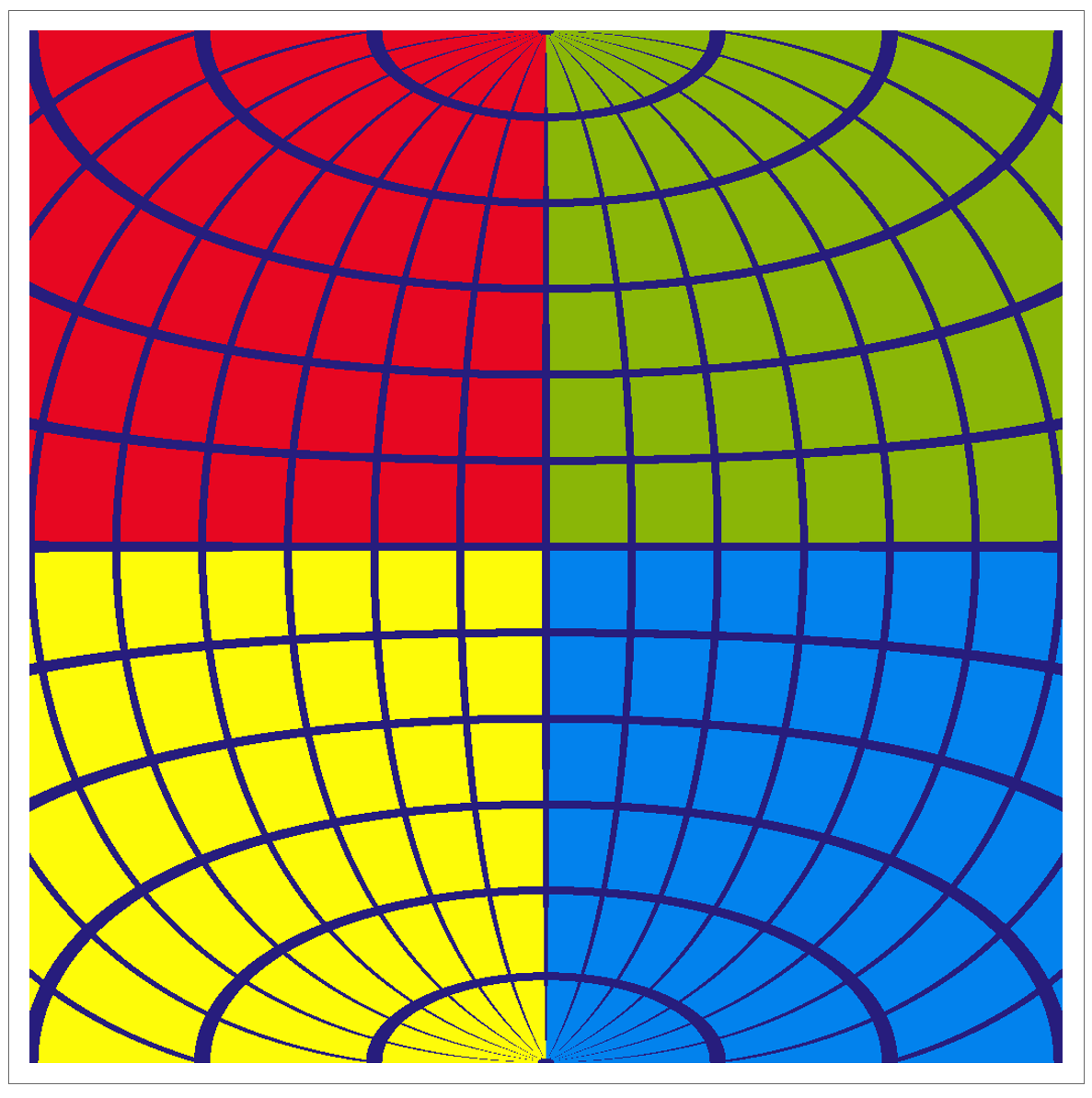}
\caption{
Left: The setup of the observer and the coordinate system.
Right: The image of the celestial sky in the Minkowski spacetime.}
\label{fig:image-setup}
\end{figure}

The image of the black hole spacetime can be obtained numerically using the backward ray tracing method.
It is essentially a technique to numerically integrate the trajectories of individual photons that are emanated from the observer, giving the effective potentials.
The method is based on the fact that a null geodesic will end up either at a pixel on the celestial sky, which will be subsequently assigned to a given color as defined beforehand, or at the outer event horizon of the black hole, which will be denoted as a {\it black} dot.
Such a strategy is feasible as long as the metric is stationary because any geodesic is reversible in time in such a spacetime configuration. 
By enumerating and painting the entire solid angle of the observer, one can depict the image of the black hole spacetime from the observer's perspective.

As shown in Fig.~\ref{fig:image-setup}, one places the black hole at the origin of the coordinate system and the observer on the equatorial plane.
A null geodesic emanated from the observer is governed by initial conditions or the conserved quantities.
Specifically, they can be expressed either in terms of the solid angle $(\theta_O, \phi_O)$ in the observer's local sky or the conserved quantities. 
In either case, they are determined by two free parameters.
The initial angular direction of the geodesic determines the coordinates of the corresponding pixel.
The color of the pixel is governed by the eventual fate of the photon, as it either escapes to spatial infinity or captured by the black hole event horizon.
For the latter, it corresponds to a black pixel.
For former, the eventual angular direction when the geodesic reaches spatial infinity determine the pixel's color.
The conserved quantities can be used readily in the equation of motion Eq.~\eqref{eosGeodesic} to derive the trajectory and, subsequently the final angular motion of the photon.

One needs to choose the basis vectors in the observer's local inertial frame, and two transforms are involved.
The first one is a coordinate transform that turns the angular directions $(\theta, \phi)$ in the observer's local sky into the two-dimensional coordinates $(x, y)$ utilized in depicting the resultant black hole image, which reads
\bqn
\theta &=& \arccos\left(\cos\left(\frac{\pi}{4} y\right) \cos\left(\frac{\pi}{4} x\right)\right),\nb \\
\phi &=& \arctan\left(\frac{-\tan(\frac{\pi}{4} y)}{\sin(\frac{\pi}{4} x)}\right) .
\eqn
The above transform is somewhat arbitrary, as one needs to define a map from a $S^2$ surface to a flat sheet.

The second one evaluates the covariant four-momentum of the photon using the energy and the three-momentum of the photon in the observer's local sky.
This is essentially governed by the choice of the basis vectors in the observer's local inertial frame.
Following Ref.~\cite{Cunha:2016bpi,wang2018shadows,Johannsen:2013vgc}, we have
\bqn
p_t &=& -\sqrt{-\frac{\Delta}{g_{\phi\phi}}}-\sin\theta\cos\phi \frac{g_{t\phi}}{\sqrt{g_{\phi\phi}}},\nb \\
p_r &=& \cos\theta \sqrt{g_{rr}},\nb \\
p_\theta &=& \sin\theta\sin\phi  \sqrt{g_{\theta\theta}},\nb \\
p_\phi &=& \sin\theta\cos\phi  \sqrt{g_{\phi\phi}},
\eqn
where
\bqn
\Delta &=& g_{tt} g_{\phi \phi}-g_{t\phi}^2 .
\eqn
In the above expressions, we have assumed that all the emitted photons have unit energy in the local inertial frame.


\begin{figure}[htb]
\includegraphics[scale=0.34]{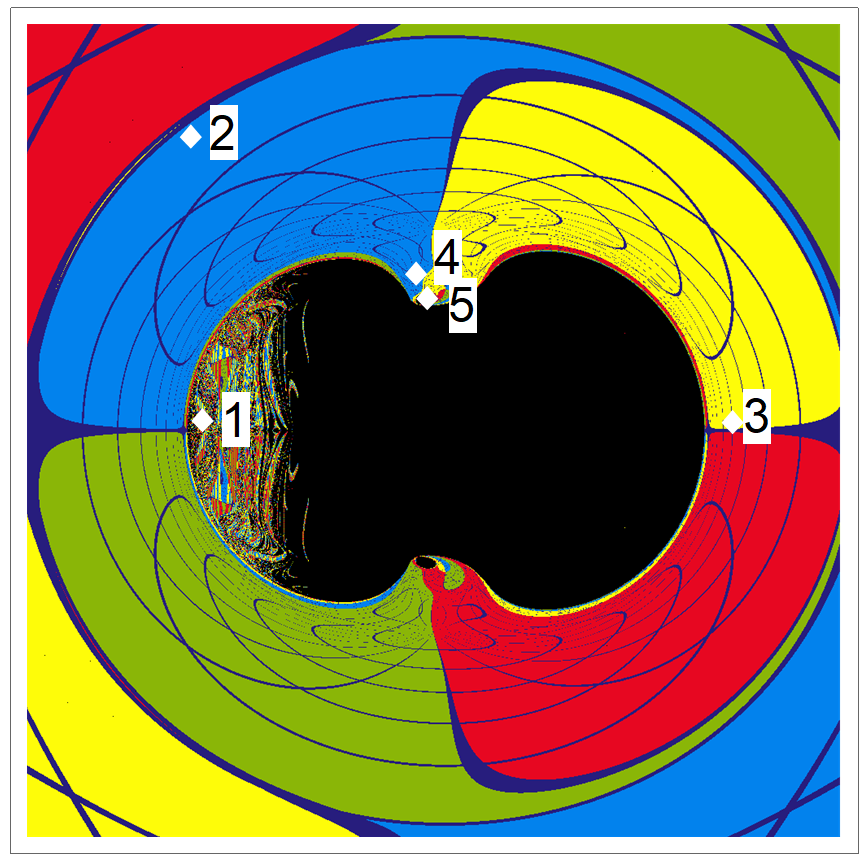}
\includegraphics[scale=0.27]{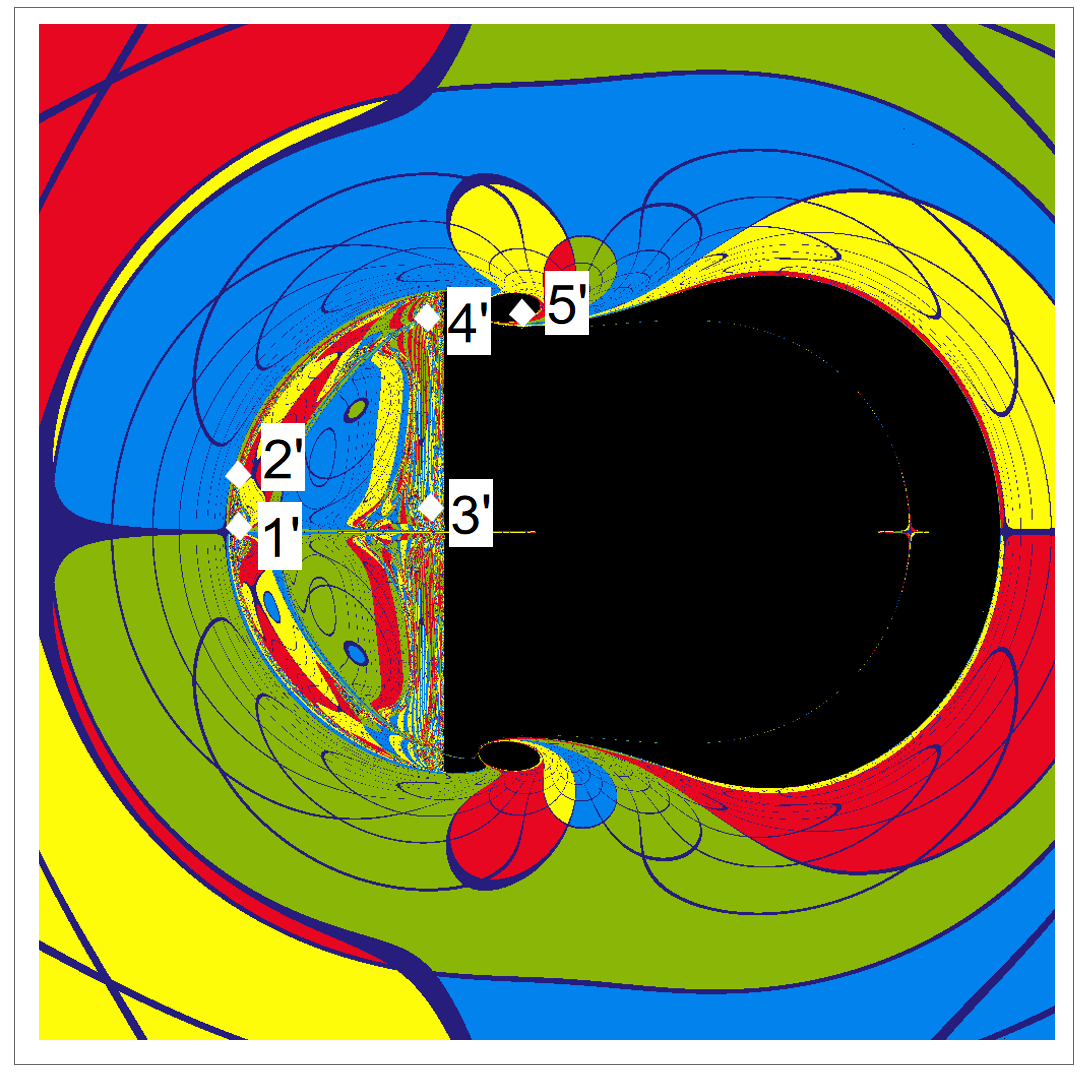}
\caption{
The images of the Manko-Noviko black hole spacetime using different metric parameters.
The calculations are carried out using the metric parameter set 1 (left) and 2 (right), specified in Eq.~\eqref{metricPar}.}
\label{fig:scenario_selection}
\end{figure}

The resultant black hole images are presented in Fig.~\ref{fig:scenario_selection}, where chaotic lensing is observed by the edge of the central black hole shadow in both cases, particularly those on the left-hand side.
In what follows, we elaborate on further analysis of the emergence of chaotic lensing and its connection with the properties of geodesic and effective potentials.

\subsection{Connection between pockets and chaotic lensing}

It was first pointed out by Cunha P V P, Grover J, Herdeiro C, et al.{\it et al.}~\cite{Cunha2016chaotic} that it is intuitive to observe that the formation of the pocket discussed in the last section might be related to the black hole shadow.
As denoted in Fig.~\ref{fig:scenario_selection}, we will explore several distinct scenarios, particularly on the edge of the image's chaotic region, to study the key elements that lead to such a phenomenon.

\begin{figure}[htb]
\includegraphics[scale=0.27]{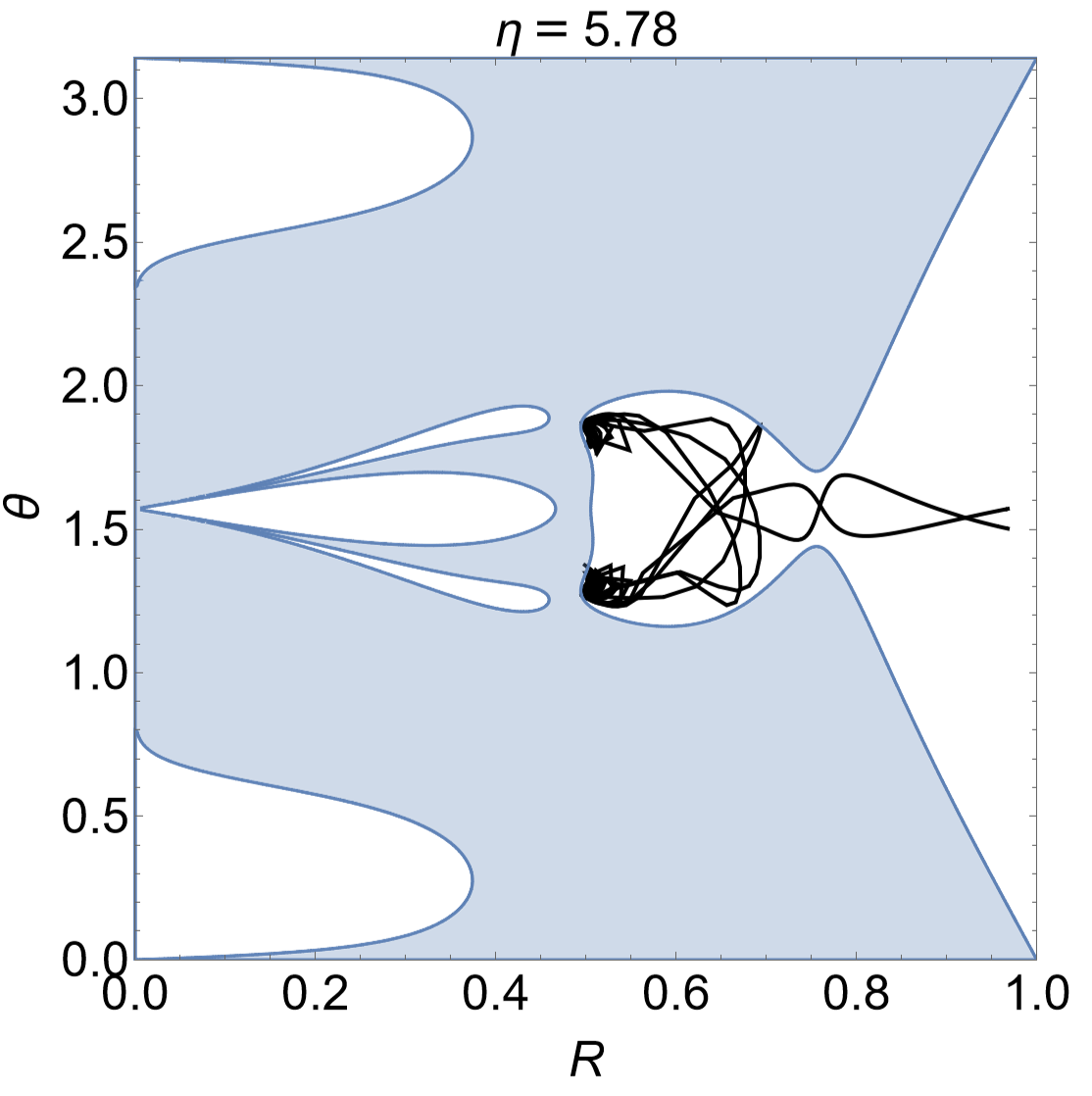}
\includegraphics[scale=0.26]{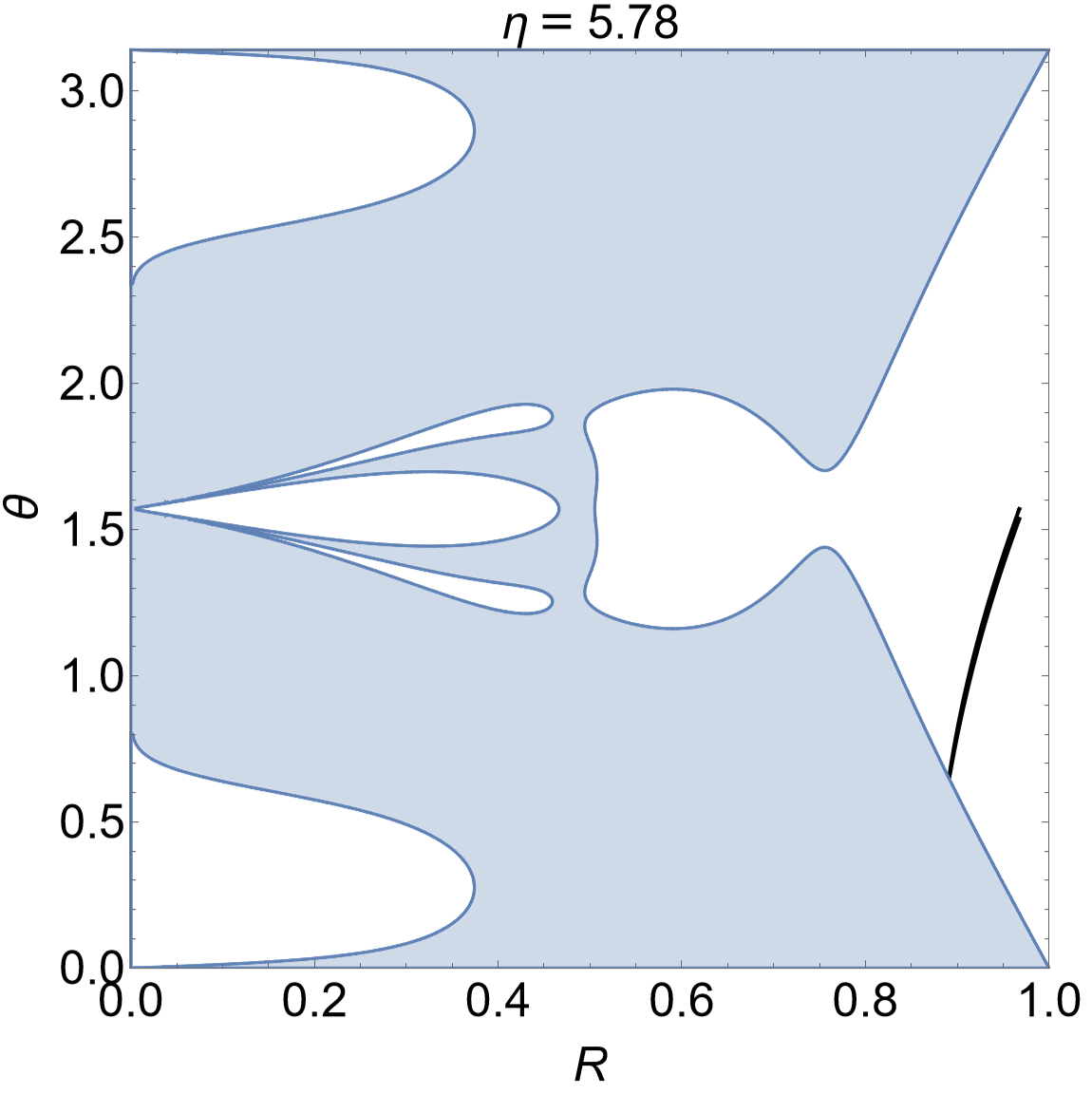}
\includegraphics[scale=0.26]{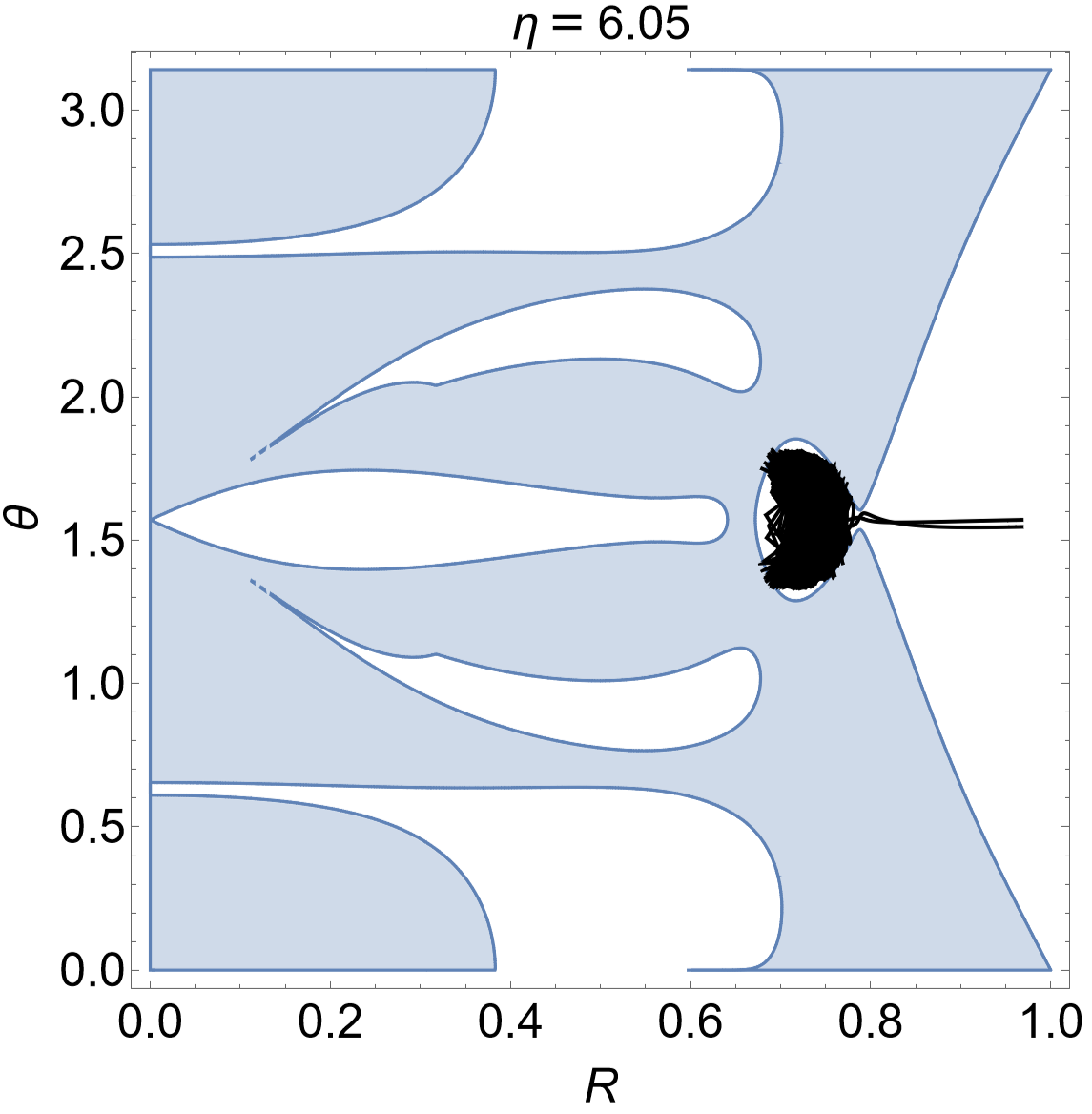}
\includegraphics[scale=0.26]{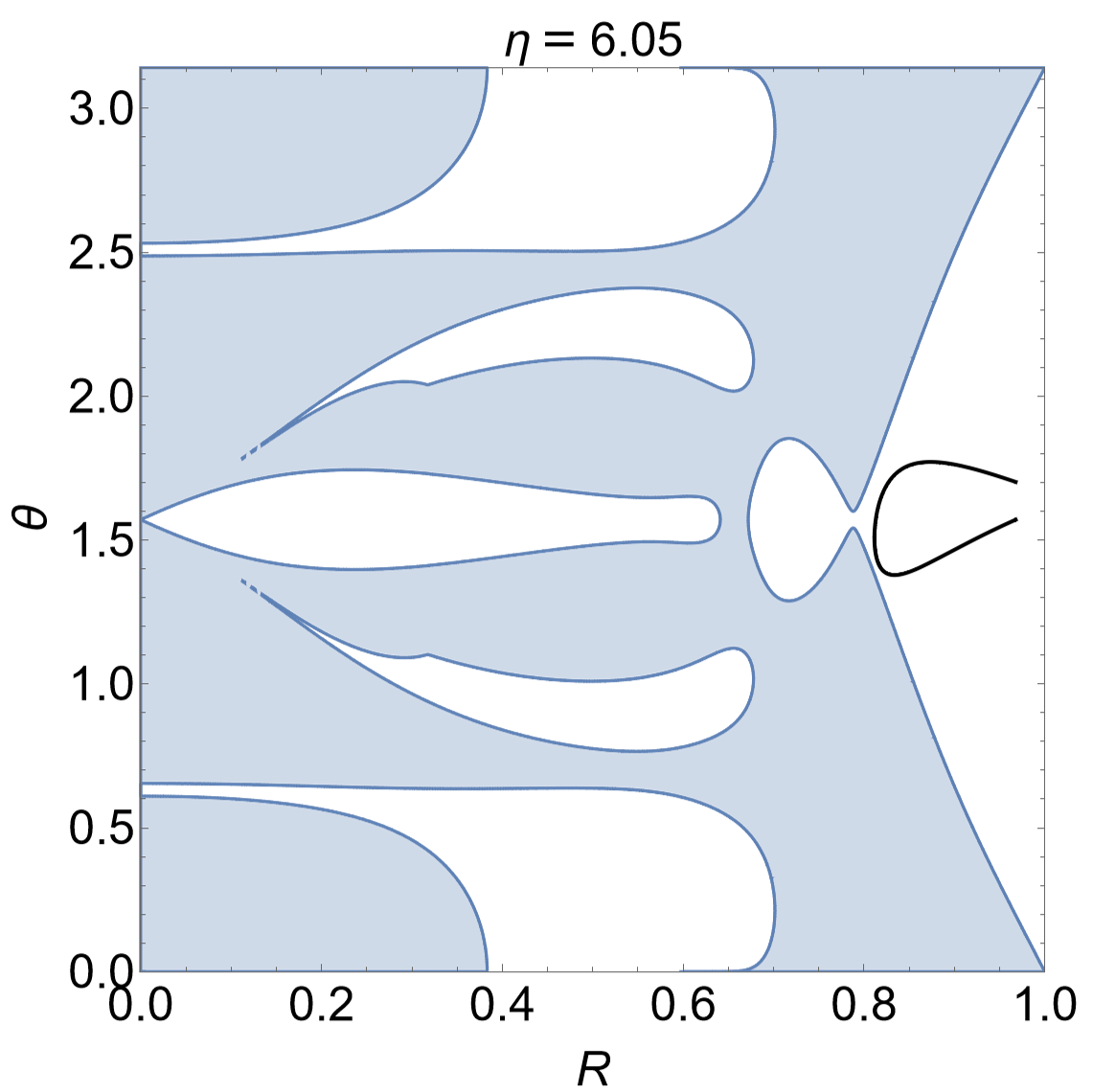}
\caption{The illustrations for the first scenario: 
two sets of trajectories of adjacent null geodesics with the presence of a pocket in the effective potential.}
\label{fig:pocket_geo_traj}
\end{figure}

Three different scenarios will be elaborated.
We first consider a physically straightforward scenario with the presence of a pocket.
As discussed above, a pocket might be formed for a given $\eta > 0$ owing to the contours of $h_-$.
In this context, we consider two specific cases, as shown in the top row of Fig.~\ref{fig:pocket_geo_traj}, with $\eta=5.78$.
The calculations are performed using the first set of metric parameters in Eq.~\eqref{metricPar} and denoted in the left panel of Fig.~\eqref{fig:scenario_selection} by the labels $1$ and $2$.
Since the pocket implies an area of accessible spacetime region which a narrow opening, it indicates that once a geodesic enters the area, it will keep bouncing against the wall of the pocket repeatedly until it eventually escapes to infinity through a narrow opening.
However, owing to the size of the opening, the number of bounces is typically significant.
In other words, any insignificant deviation in its initial state will become significantly amplified, and therefore it is increasingly sensitive to the initial state, namely, the incident angle of the geodesic.
Meanwhile, each time the number of bouncing increases by a unit, it implies that the geodesic's resultant solid angle subsequently involves by an entire period in terms of either one of the two angular coordinates. 
Therefore, the pixel's color varies significantly in an insignificant coordinate interval, giving rise to chaotic lensing.
This is intuitive to understand due to the strongly nonlinear nature of a deterministic system, demonstrated by the geodesic equation of motion.

A typical geodesic trajectory of the above case is shown on the top left plot of Fig.~\ref{fig:pocket_geo_traj}.
In this case, once a geodesic enters the pocket, the emergence of chaotic lensing is triggered.
In the right panel of Fig.~\ref{fig:pocket_geo_traj}, we show an example that the geodesic does not enter the pocket but bounces back immediately at the outer wall of the accessible region.
For the latter case, it is apparent that one will not observe chaotic lensing, even though a pocket is present.
The geodesic corresponds to a colored pixel on the background celestial sky.
Two examples of this case are presented on the top right plot of Fig.~\ref{fig:pocket_geo_traj}.
As observed, the two geodesics on the top row of Fig.~\ref{fig:pocket_geo_traj} correspond to pixels adjacent in the resultant black hole image. 
One is located inside the chaotic region, while the other stays on the immediate outside. 
The two geodesics on the bottom row possess similar features.
They are obtained using the second set of metric parameters and denoted by $1'$ and $2'$ in the right panel of Fig.~\ref{fig:scenario_selection}.
It is also worth pointing out that the geodesic cannot enter the horizon of the black hole.
As shown in Fig.~\ref{fig:pocket_geo_traj}, any possible trajectory to the interior of the black hole is completely walled off by the effective potentials.
Therefore, it always corresponds to a colored pixel, either chaotic or not.

\begin{figure}[htb]
\includegraphics[scale=0.27]{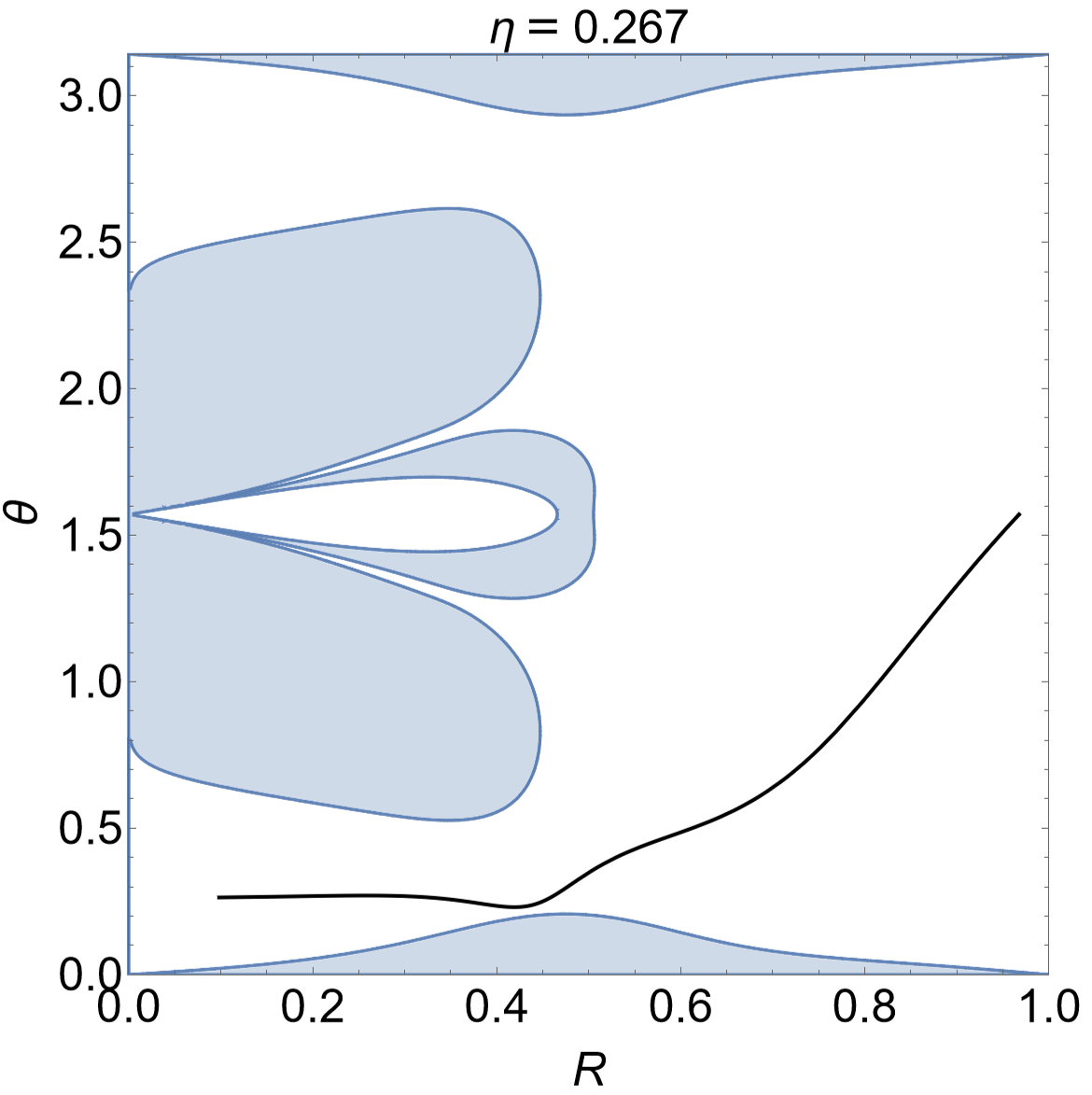}
\includegraphics[scale=0.31]{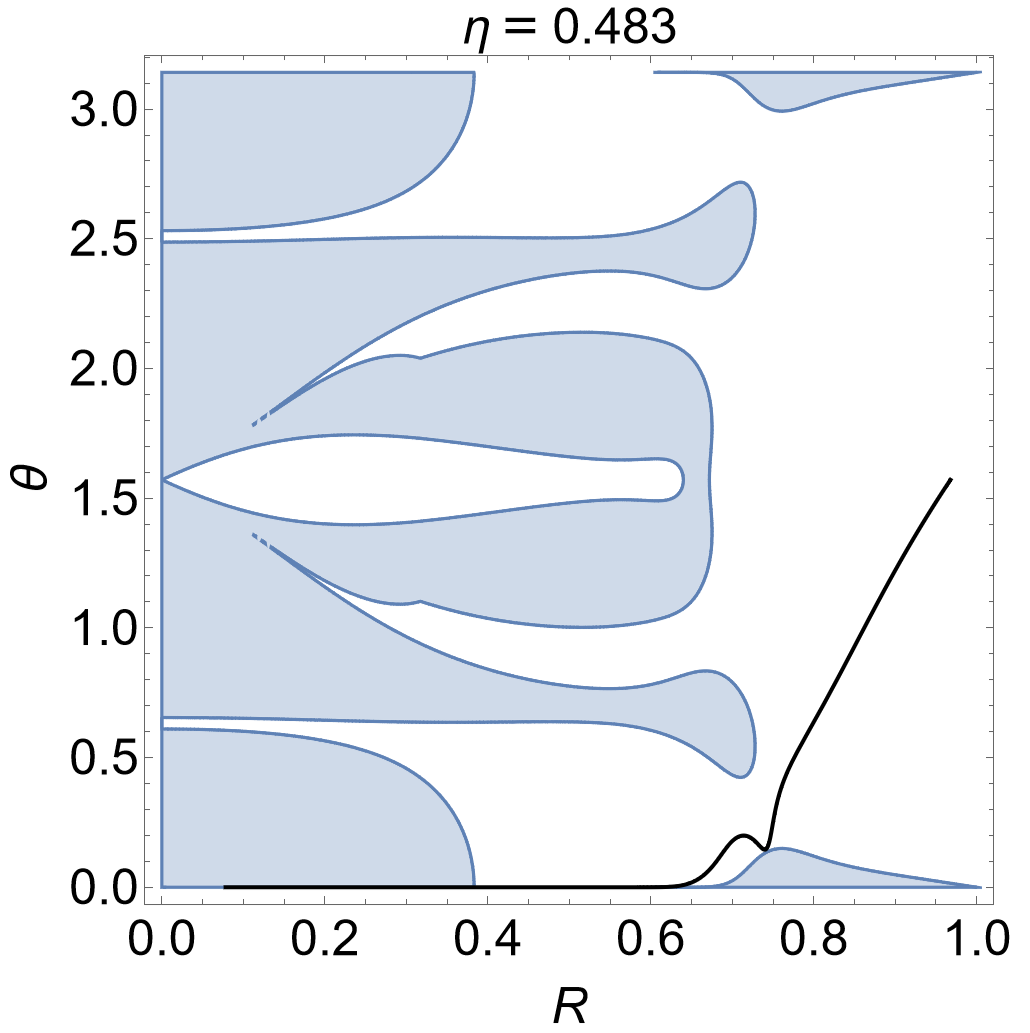}
\caption{
The illustrations for the second scenario:
the trajectories of null geodesics which fall directly into the outer event horizon without the presence of a pocket in the effective potential.}
\label{fig:pocket_geo_traj2}
\end{figure}

The second scenario discussed here is also intuitive.
As shown in Fig.~\ref{fig:pocket_geo_traj2}, the effective potential possesses a gap at a certain zenith angle.
As a result, a geodesic may fall directly into the horizon of the black hole.
Such a case is illustrated by the geodesic shown in Fig.~\ref{fig:pocket_geo_traj2}.
They are obtained using the two sets of metric parameters and denoted by the labels $5$ and $5'$ in the left and right panels of Fig.~\ref{fig:scenario_selection}.
Indeed, geodesic may also bounce off from the effective potential in this case, similar to the right column shown in Fig.~\ref{fig:pocket_geo_traj}.

\begin{figure}[htb]
\includegraphics[scale=0.26]{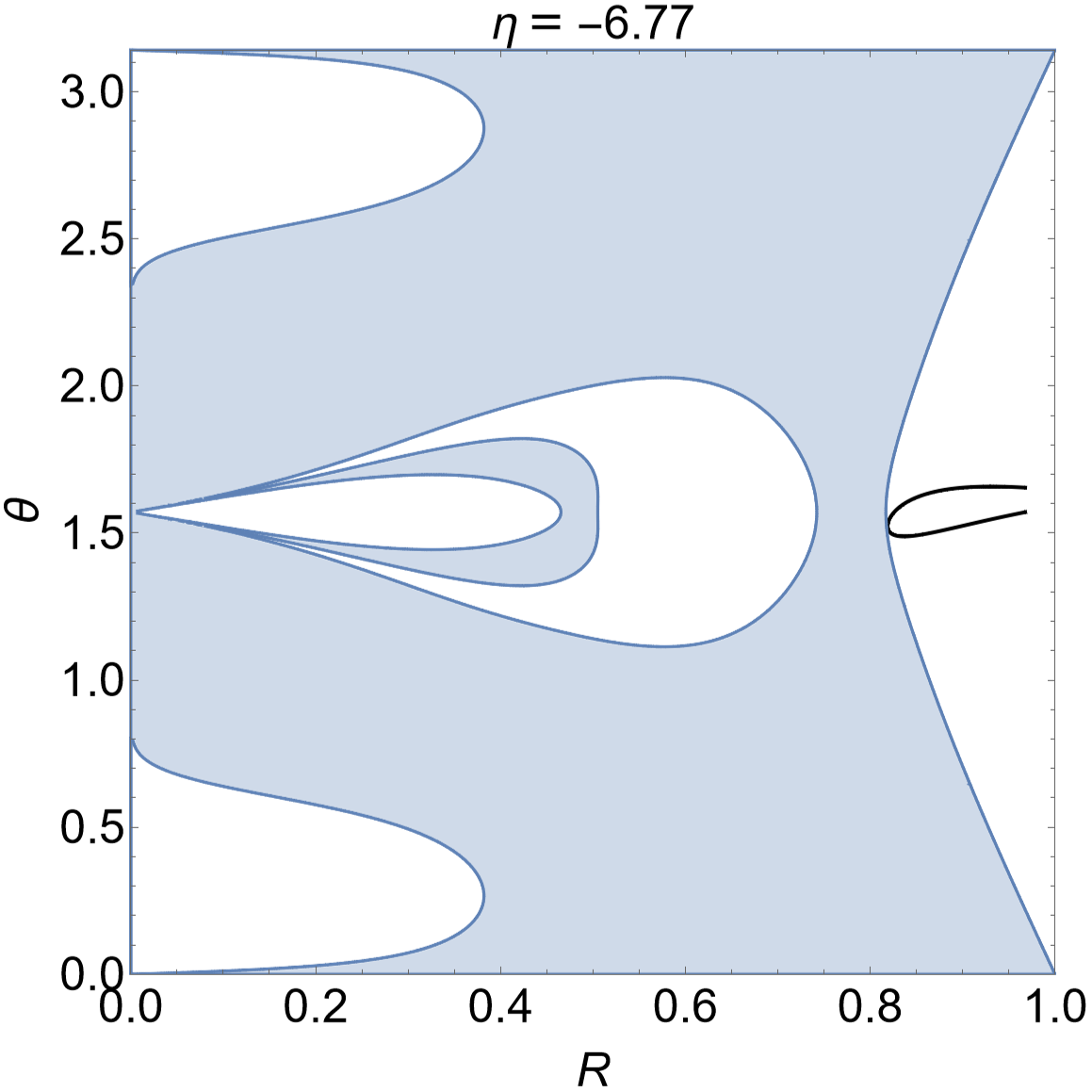}
\includegraphics[scale=0.26]{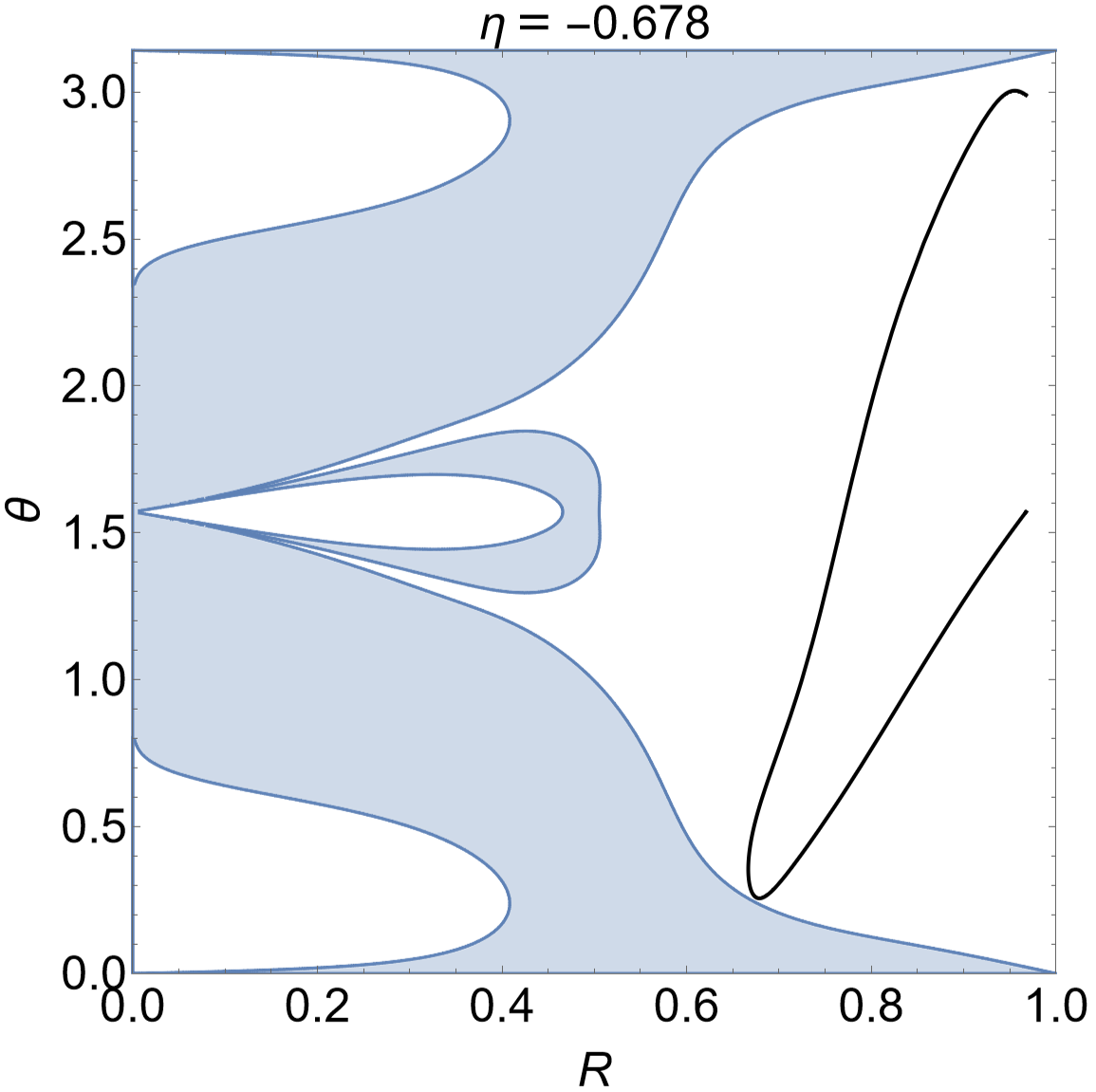}
\includegraphics[scale=0.285]{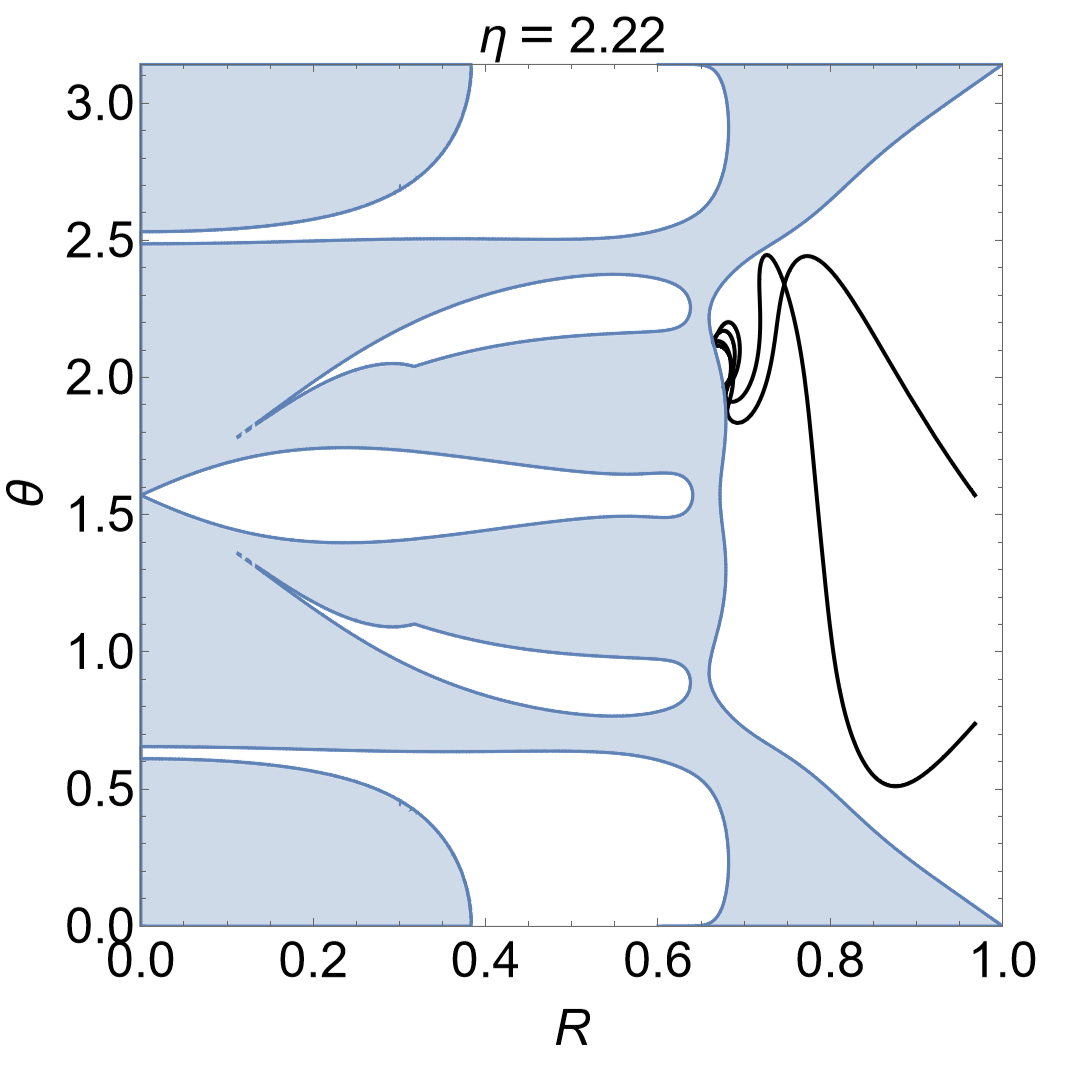}
\includegraphics[scale=0.26]{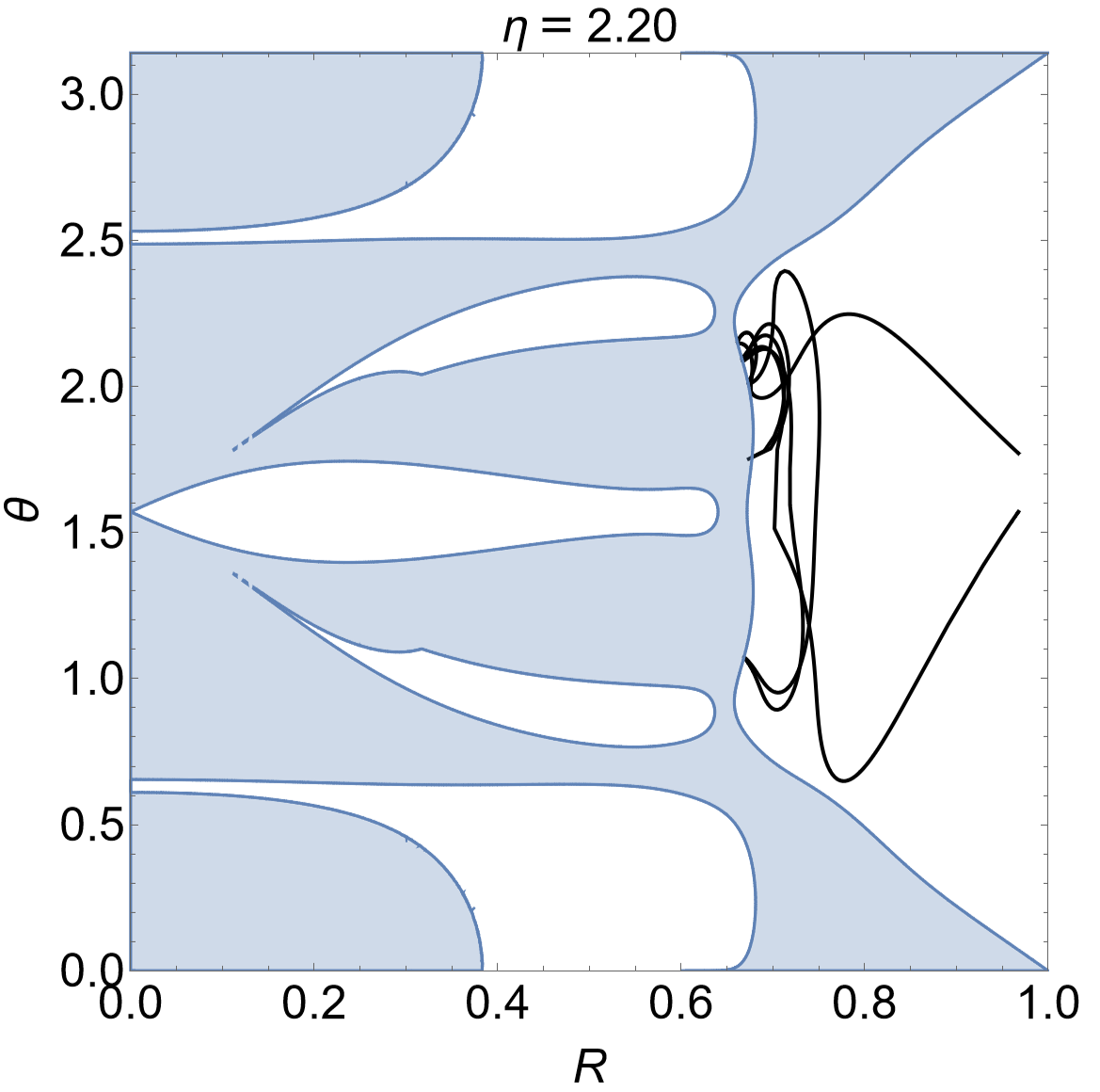}
\caption{ 
The illustrations for the third scenario: 
two sets of trajectories of null geodesics without an unambiguous presence of a pocket in the effective potential.}
\label{fig:pocket_geo_traj3}
\end{figure}

The third and last scenario corresponds to the emergence of chaotic lensing without forming any pocket.
While the effective potentials entirely block the geodesic's access to the horizon.  
Intuitively, in such a scenario, chaotic lensing is not expected.
This is illustrated by the two geodesics on the top row of Fig.~\ref{fig:pocket_geo_traj3}.
They are obtained using the first set of metric parameters and denoted by the labels $3'$ and $4'$ in the right panel of Fig.~\ref{fig:scenario_selection}.
However, as it turns out, this is not necessarily the only possibility.
On the bottom row of Fig.~\ref{fig:pocket_geo_traj3}, chaotic lensing is observed for the two geodesics for black hole metric using the second set of metric parameters.
This is in accordance with their locations denoted on the black hole image shown in the right plot of Fig.~\ref{fig:scenario_selection}.
It is somehow not straightforward to judge whether a pocket (if any, with a relatively wide opening) is formed.
Nonetheless, the geodesic in question seems to be attracted to the bound of the accessible region and bounced back.
The above discussions indicate that the pocket alone, although an important asset, might not be enough to determine the emergence of chaotic lensing.
In the following subsection, we explore another intriguing factor that involves the photons' kinematics: the turning points of the null geodesic.

\subsection{Connection with angular and radial accelerations}

In this subsection, we turn to discuss the impact of angular and radial accelerations of the null geodesics.
Specifically, we explore their distributions at the turning points and their relation with chaotic lensing.
Such a consideration is motivated by the observation that the trajectory of null geodesics tends to converge in specific regions of spacetime.
For instance, when chaotic lensing emerges as they are ``trapped'' for an extensive period inside a pocket, the collection of the trajectory of the null geodesic are not necessarily reminiscent of that of an ergodic process. 
In other words, the trajectory may not occupy the whole chamber, namely, the entire coordinate space inside the pocket.
Such a feature can be primarily understood by investigating the accelerations at the turning points in the trajectory.

The turning points on the null geodesic are defined intuitively as follows~\cite{Cunha2016chaotic}.
A turning point in the radial direction implies $\dot{r}=0$, while that in the zenith direction means $\dot{\theta}=0$.
By making use of the geodesic equation Eq.~\eqref{eosGeodesic}, it is not difficult to find the accelerations that satisfy 
\begin{equation}
\dot{p}_r=-\frac{1}{2 E^2}(V\partial_r \ln [g_{\theta \theta}]+\partial_r V), \label{accRadial}
\end{equation}
and
\begin{equation}
\dot{p}_\theta=-\frac{1}{2 E^2}(V\partial_\theta \ln [g_{r r}]+\partial_\theta V) . \label{accAngular}
\end{equation}
Here Eq.~\eqref{accRadial} gives the radial acceleration at the radial turning point, while Eq.~\eqref{accAngular} is the zenith acceleration at the angular turning point.
These two equations furnish a distribution of accelerations as a function of spacetime coordinates.

Besides the specific shape of the accessible region, such as a pocket, it is also possible to constrain the motion of a geodesic inside a specific spacetime region delimited by significant accelerations with opposite signs.
Intuitively, acceleration furnishes a {\it kinematic} constraint on the motion of geodesic, while a pocket enforces a {\it static} bound by solid walls around the area. 

\begin{figure}[htb]
\includegraphics[scale=0.256]{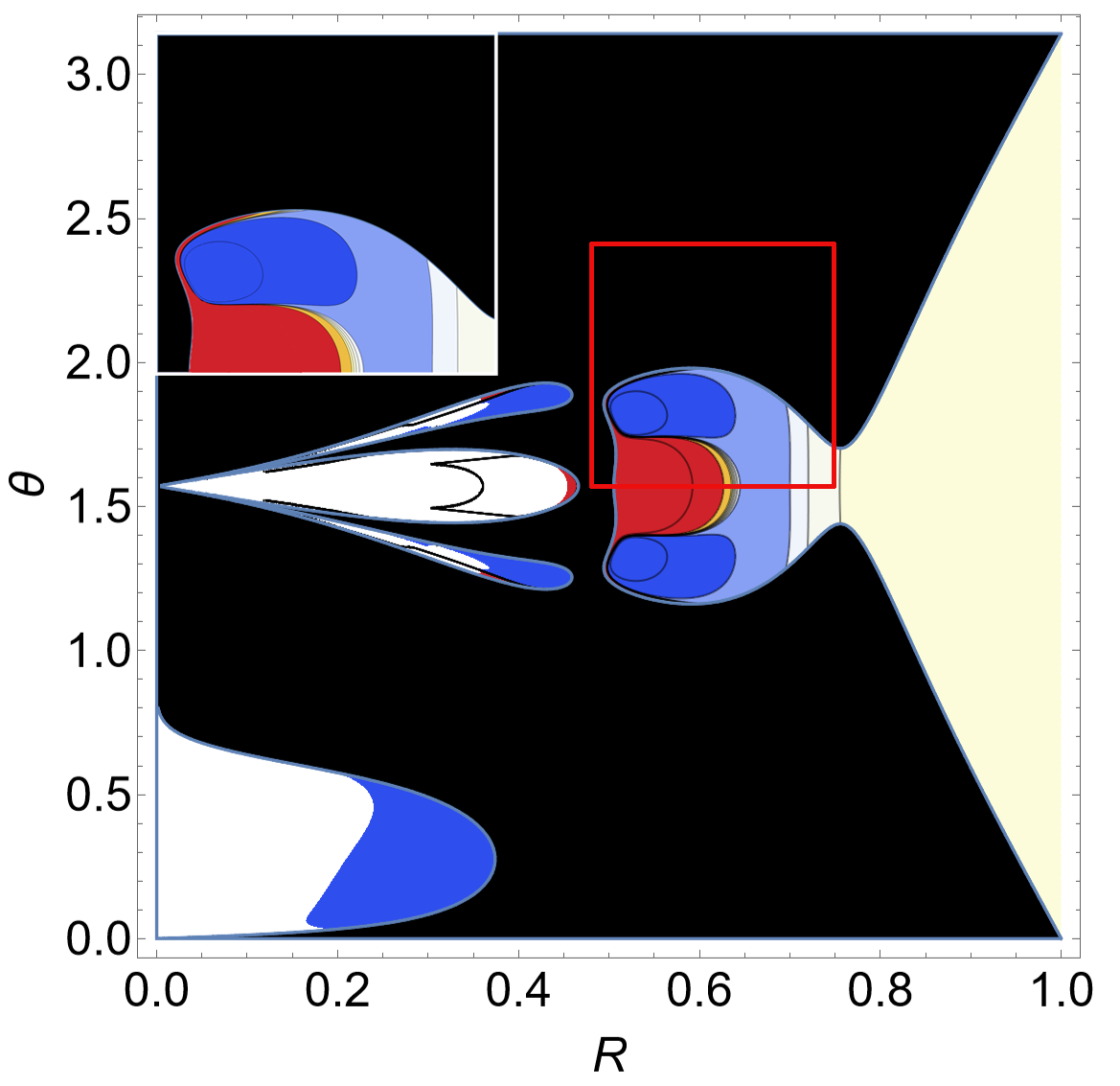}
\includegraphics[scale=0.25]{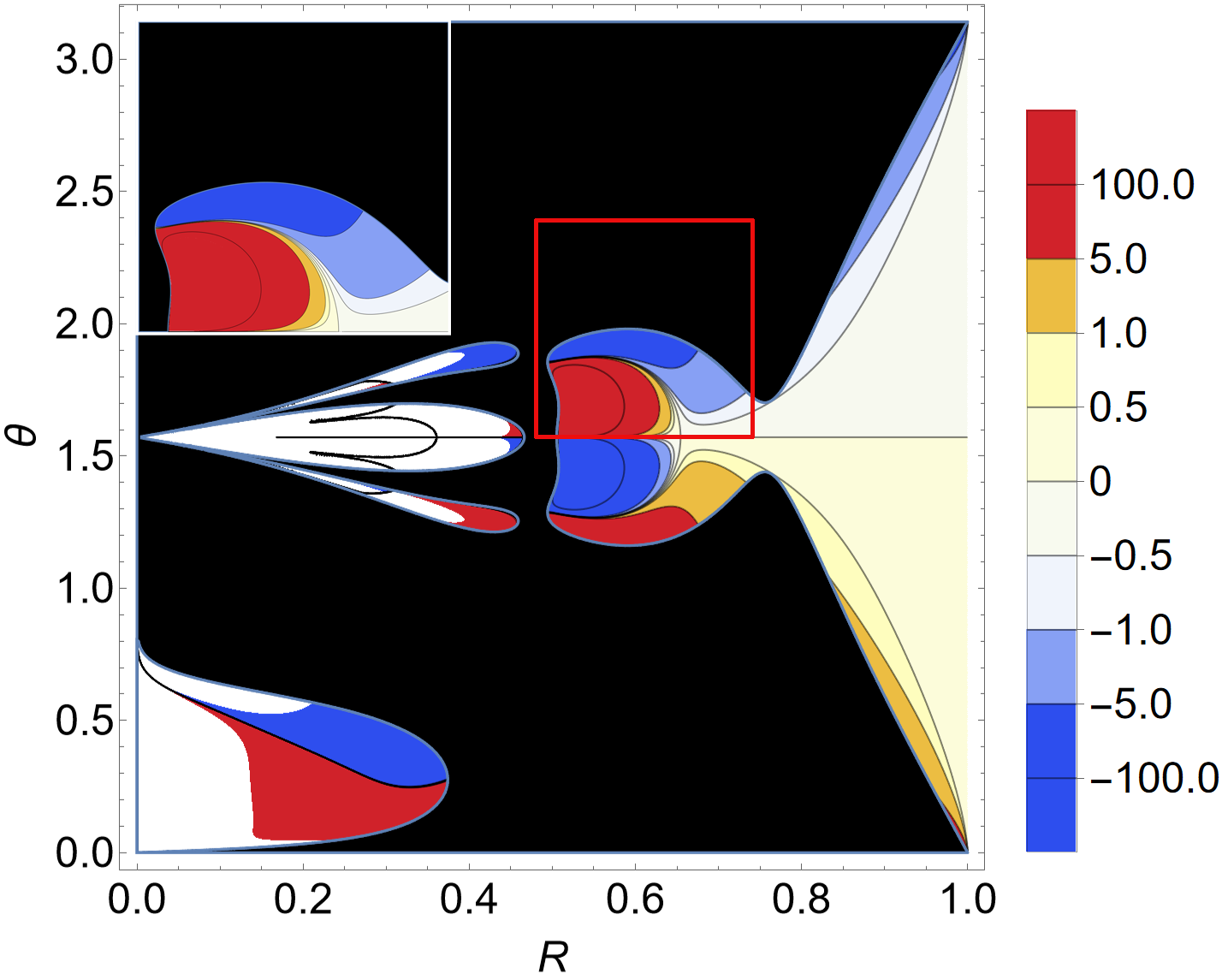}
\caption{The distributions of the radial $\dot{p}_r$ (left) and zenith $\dot{p}_\theta$ (right) accelerations on top of the effective potential with a pocket.}
\label{fig:accRadialAngular}
\end{figure}

\begin{figure}[htb]
\includegraphics[scale=0.268]{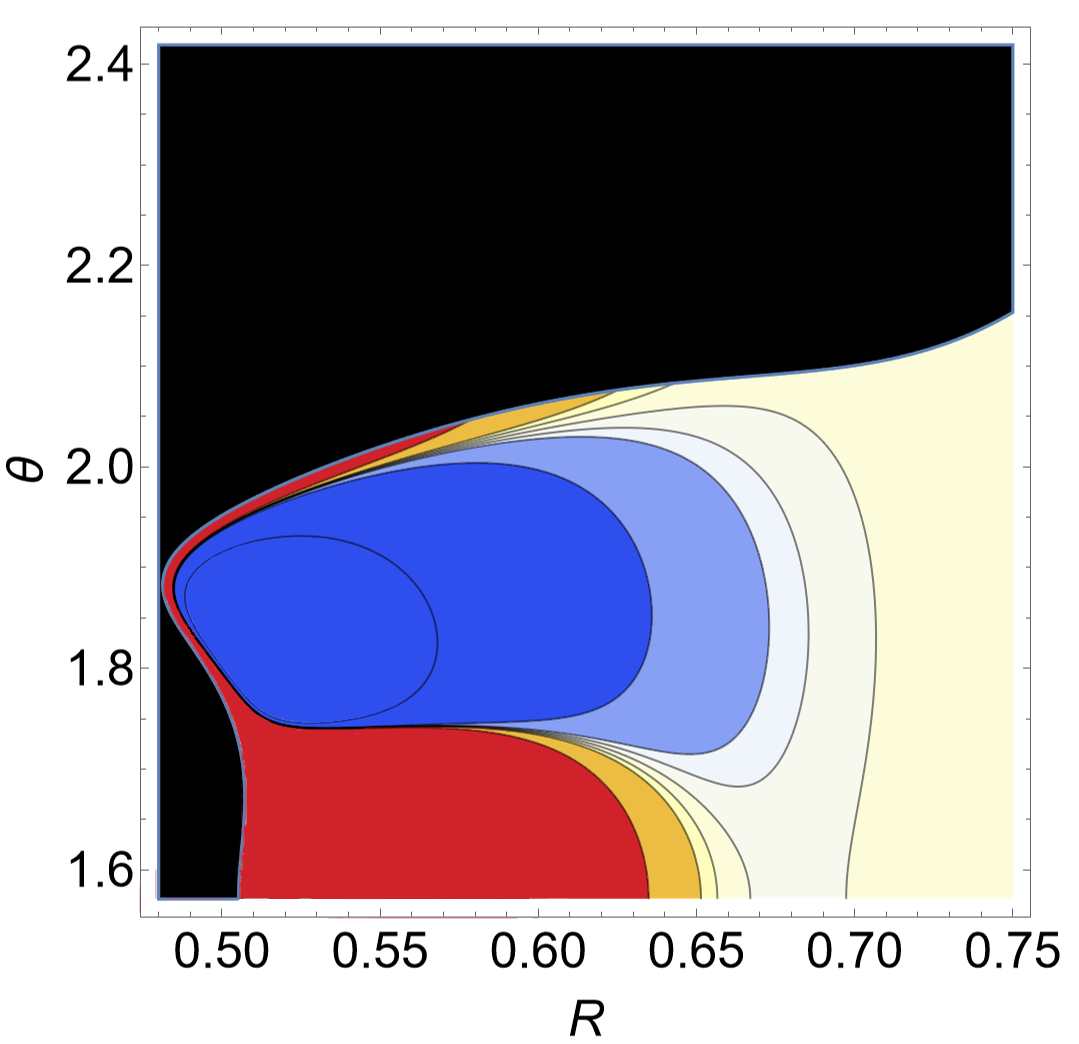}
\includegraphics[scale=0.27]{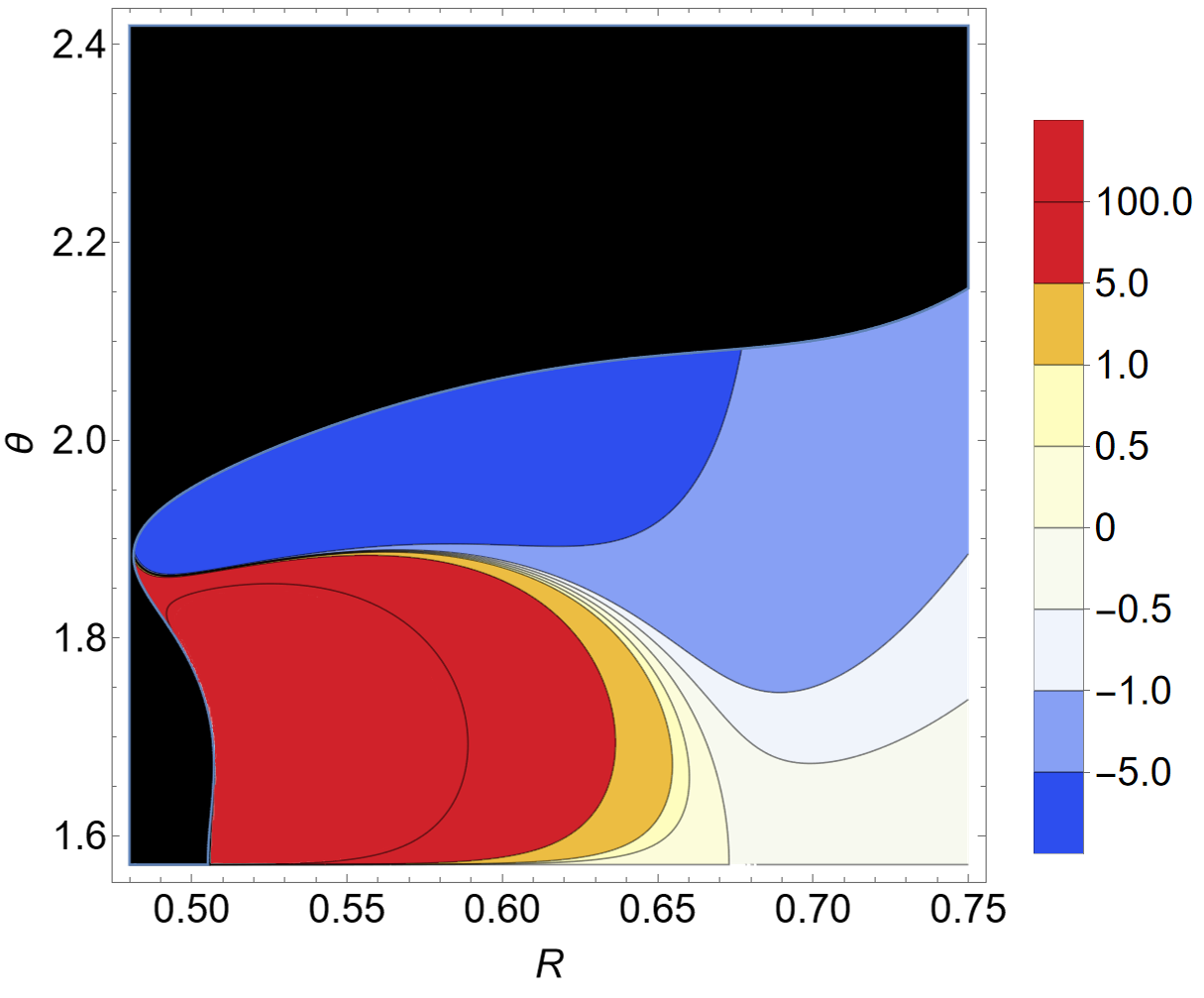}
\caption{The distributions of the radial $\dot{p}_r$ (left) and zenith $\dot{p}_\theta$ (right) accelerations on top of the effective potential without a pocket.}
\label{fig:accRadialAngular2}
\end{figure}

The results are shown in Fig.~\ref{fig:accRadialAngular} and~\ref{fig:accRadialAngular2} using the first set of metric parameters.
In the left panel of Fig.~\ref{fig:accRadialAngular}, one observes that the radial acceleration is positive at the pocket's bottom (c.f., the zoomed-in area shown in the upper-left corner).
It is in accordance with the fact that as the photon approaches the bottom of the pocket, it will get bounced off by the wall.
Around the opening of the pocket, on the other hand, there is no significant acceleration, indicating that the photon can enter the pocket most freely once it finds an appropriate incident angle.
However, it is notable that, outside the opening, there is a region of positive radial acceleration (not shown in the plot).
The latter implies that a photon will likely get deflected off in this area.
It is also observed that the distribution is symmetric with respect to the symmetric axis of the pocket, $\theta=\pi/2$.

The distribution of the zenith acceleration shown in the right panel of Fig.~\ref{fig:accRadialAngular} is also in accordance with the shape of the pocket.
Different from the radial acceleration, it is observed that the distribution is antisymmetric with respect to the symmetric axis of the pocket, $\theta=\pi/2$.
Two narrow regions of negative and positive accelerations are located at the upper and lower edges of the pocket.
These two regions prevent the photon from traversing the bound of the pocket.
On the other hand, the angular acceleration in the pocket's upper half is mostly positive, while that in the lower half is primarily negative.
It indicates that the photon is constantly pushed off from the center of the chamber, consistent with the previous observations that the null geodesic is mainly concentrated in a narrow region inside the pocket.

Based on the above discussions, one concludes that the kinematic constraint effectively provides a barrier that is most consistent with the static constraint furnished by the accessible region. 
Now, to investigate the effect of kinematic constraint alone, we turn to study the distribution of the acceleration where the pocket does not form.
The results are shown in Fig.~\ref{fig:accRadialAngular2}.
In the left panel of Fig.~\ref{fig:accRadialAngular2}, a robust potential barrier (shown in blue) in the radial direction is observed, even though the effective potential is wide open in the area.
In the angular direction, a similar feature is observed.
The above kinematic constraint is consistent with the chaotic lensing observed above in the bottom row of Fig.~\ref{fig:pocket_geo_traj3}, where a pocket is absent.

\section{Concluding remarks}\label{section4}

To summarize, in this work, we explored the properties and emergence of chaotic gravitational lensing in the Manko-Novikov black hole spacetimes. 
To explore the underlying physics, we analyzed the boundaries of the accessible region in terms of the contours of the effective potentials $h_\pm$ and the ergoregions.
An interplay between the ergoregion and effective potential defines the static bound of the photon's accessible region.
Chaotic lensing typically occurs with the presence of a pocket in the effective potential.
On the other hand, it was observed that the above criterion is insufficient.
In this regard, we further studied the angular and radial acceleration at the turning points of geodesics.
We argue that the latter furnishes a kinematic constraint for geodesics.
Combining the above two criteria reasonably explains the observed chaotic lensing in the Manko-Novikov black hole spacetime. 
It is therefore concluded that the onset of the chaotic lensing is crucially related to the static and kinematic constraints derived from the effective potential and the radial and angular accelerations.

In literature, chaotic lensing is an intriguing phenomenon that has only been discovered for a few metrics~\cite{Cunha2016chaotic, wang2018shadows, wang2018chaotic}.
However, an unambiguous criterion for the emergence of such a phenomenon is yet to be established.
The present study involved an attempt to explore the relevant factors in detail regarding static and kinematic constraints.
Under the circumstance of such constraints, an arbitrarily small deviation in the trajectory of the incident photon will be significantly amplified through an extensive evolution in the spacetime region.
As a result, the chaotic phenomenon in strong gravitational lensing demonstrates the complexity of the highly nonlinear deterministic gravitational system, and intriguing implications are entailed.
It is also curious to point out that chaotic systems can be successfully analyzed and, to a certain degree, predicted by machine learning algorithm~\cite{pathak2018model}.
It is, therefore, intriguing whether such an approach might be adopted for the gravitational system.

\section*{Acknowledgements}
We are thankful for the insightful discussions with Songbai Chen and Weisheng Huang.
This work is supported by the National Natural Science Foundation of China (NNSFC) under Grants No. 12005077 and Guangdong Basic and Applied Basic Research Foundation under Grant No. 2021A1515012374.
We also gratefully acknowledge the financial support from Brazilian agencies 
Funda\c{c}\~ao de Amparo \`a Pesquisa do Estado de S\~ao Paulo (FAPESP), 
Funda\c{c}\~ao de Amparo \`a Pesquisa do Estado do Rio de Janeiro (FAPERJ), 
Conselho Nacional de Desenvolvimento Cient\'{\i}fico e Tecnol\'ogico (CNPq), 
and Coordena\c{c}\~ao de Aperfei\c{c}oamento de Pessoal de N\'ivel Superior (CAPES).
A part of this work was developed under the project Institutos Nacionais de Ciências e Tecnologia - Física Nuclear e Aplicações (INCT/FNA) Proc. No. 464898/2014-5.
This research is also supported by the Center for Scientific Computing (NCC/GridUNESP) of S\~ao Paulo State University (UNESP).
\bibliographystyle{h-physrev}
\bibliography{ref}
 
\end{document}